\documentclass{aastex63}
\usepackage{graphicx}
\usepackage{subfigure}
\usepackage{multirow}
\usepackage{comment}
\usepackage{natbib}
\usepackage{mathtools}
\usepackage{mathrsfs}
\usepackage{fontenc}
\usepackage{color}
\usepackage{url}
\usepackage{hyperref}
\usepackage{gensymb}
\usepackage{pifont}
\newcommand{\lbol}{$L_{\rm bol}$}

\newcommand{\lum}{\rm erg~s$^{-1}$}

\newcommand       \Angstrom     {\,{\rm \AA}}

\newcommand       \g            {\,{\rm g}}
\newcommand       \K            {\,{\rm K}}

\newcommand       \s            {\,{\rm s}}

\newcommand       \gtsim        {\gtrsim}

\newcommand       \mum          {{\rm \mu m}}

\newcommand       \usfr         {{M_\odot}\,\rm yr^{-1}}

\newcommand{\etal}{\textrm{et al.\ }}

\newcommand{\eg}{\textrm{e.g., }}

\newcommand{\agnfrac}{ f_{\mathrm {AGN}} }
\newcommand{\agnfracpg}{ f^{\rm PG}_{\rm AGN} }

\newcommand{\sfrir}{SFR$_{\rm IR}$}
\newcommand{\sfrpah}{SFR$_{\rm PAH}$}
\newcommand{\sfrneon}{SFR$_{\rm Ne}$}

\submitjournal{ApJ}

\begin{document}

\title{The Ionization and Destruction of Polycyclic Aromatic Hydrocarbons in Powerful Quasars}


\author{Yanxia Xie}
\affil{Kavli Institute for Astronomy and Astrophysics, Peking University,
Beijing 100871, China}
\correspondingauthor{Yanxia Xie}
\email{ yanxia.xie@pku.edu.cn }

\author{Luis C. Ho}
\affil{Kavli Institute for Astronomy and Astrophysics, Peking University,
Beijing 100871, China}
\affil{Department of Astronomy, School of Physics, Peking University,
Beijing 100871, China}

\begin{abstract} 
We reanalyze the mid-infrared ($5-40\,\mum$) Spitzer spectra of 86 low-redshift ($z < 0.5$) Palomar-Green quasars to investigate the nature of polycyclic aromatic hydrocarbon (PAH) emission and its utility as a star formation rate (SFR) indicator for the host galaxies of luminous active galactic nuclei (AGNs).  We decompose the spectra with our recently developed template-fitting technique to measure PAH fluxes and upper limits, which we interpret using mock spectra that simulate the effects of AGN dilution.  While luminous quasars can severely dilute and affect the detectability of emission lines, PAHs are intrinsically weak in some sources that are otherwise gas-rich and vigorously forming stars, conclusively demonstrating that powerful AGNs destroy PAH molecules.  Comparing PAH-based SFRs with independent SFRs derived from the mid-infrared fine-structure neon lines and the total infrared luminosity reveals that PAHs can trace star formation activity in quasars with bolometric luminosities $\lesssim 10^{46}\, \rm erg\,s^{-1}$, but increasingly underestimate the SFR for more powerful quasars, typically by $\sim 0.5$~dex.  Relative to star-forming galaxies and low-luminosity AGNs, quasars have a comparable PAH~11.3~$\mum$/7.7~$\mum$ ratio but characteristically lower ratios of 6.2~$\mum$/7.7~$\mum$, 8.6~$\mum$/7.7~$\mum$, and 11.3~$\mum$/17.0~$\mum$.  We suggest that these trends indicate that powerful AGNs preferentially destroy small grains and enhance the PAH ionization fraction.
\end{abstract} 


\section{Introduction}\label{sec:intro} 

Originating mostly from photodissociation regions around sites where massive stars form, polycyclic aromatic hydrocarbon (PAH) emission has long been regarded as a promising indicator of star formation activity (\eg Peeters \etal 2004b; Li 2020). Effective empirical star formation rate (SFR) calibrations based on PAH emission find broad application in star-forming galaxies (SFGs) (e.g., Farrah et al. 2007; Pope et al. 2008; Treyer et al. 2010; Diamond-Stanic \& Rieke 2012; Shipley et al. 2016; Xie \& Ho 2019; Lai et al. 2020; Zhang et al. 2021), although the apparent PAH deficit in dwarf galaxies raise complications in these systems (e.g., Calzetti et al. 2007; Draine et al. 2007; Xie \& Ho 2019). PAH emission is also detected in galaxies containing active galactic nuclei (AGNs), both in their integrated mid-infrared (IR) spectra (\eg Shi \etal 2006, 2014; Hao \etal 2007; O'Dowd \etal 2009; LaMassa \etal 2012; Stierwalt \etal 2014) and from spatially resolved spectra of the central regions of nearby galaxies (\eg Mazzarella \etal 1994; Imanishi \etal 1998; Howell \etal 2007; Smith \etal 2007; Diamond-Stanic \& Rieke 2010; Gallimore \etal 2010; onz\'alez-Mart\'in \etal 2013, 2015; Sales \etal 2013; Alonso-Herrero \etal 2014, 2016, 2020; Ruschel-Dutra \etal 2014; Espara-Arredondo \etal 2018).  The prevalence of PAH emission in AGNs raises the hope that it can serve as a tool to probe star formation in active galaxies, so long as their harsh environments do not strongly affect the PAH carriers (Voit 1991, 1992). The possibility that the PAH molecules can be excited by photons from the AGN itself presents a further complication (Jensen \etal 2017).

The PAH spectra of AGNs differ from those of SFGs in several aspects. AGNs generally exhibit PAH features of lower equivalent width, most notably in the 6.2~$\mum$ and 7.7~$\mum$ bands, a characteristic widely used as a mid-IR diagnostic to identify the dominant energy source in a galaxy (\eg Genzel \etal 1998; Lutz \etal 1998; Spoon \etal 2007).  Compared to SFGs, AGNs also present lower ratios of the 6.2, 7.7, and 8.6~$\mum$ features relative to the 11.3~$\mum$ feature, to an extent that is difficult to explain in the context of dust models for star-forming regions (\eg Smith \etal 2007; O'Dowd \etal 2009; Diamond-Stanic \& Rieke 2010; Wu \etal 2010).  Smith \etal (2007) also note that the PAH~$7.7\ \mum$/$11.3\ \mum$ ratio inversely correlates with the strength of the AGN radiation field, suggesting that the hard radiation from the AGN affects the small, ionized PAH molecules that give rise to the 6--9~$\mum$ emission.  The large, neutral molecules radiating at 11.3~$\mum$ tend to be more robust against AGNs.  If so, PAH~11.3~$\mum$ may be regarded as a reliable SFR indicator for AGN host galaxies (Diamond-Stanic \& Rieke 2010, 2012; Alonso-Herrero \etal 2014).  Yet, controversy remains.  LaMassa \etal (2012) report that strong AGNs show significantly weaker PAH~11.3~$\mum$ emission and suggest that the carriers of this feature are destroyed.  Esparza-Arredondo \etal (2018) reach a similar conclusion in their analysis of the PAH~11.3~$\mum$ profile in the central few hundred parsecs of nearby active galaxies.  Additional evidence that AGNs may destroy PAHs comes from modeling the IR spectral energy distribution (SED) of luminous quasars, which indicate a mild depression of the overall PAH content with increasing quasar luminosity (Shangguan \etal 2018). 

Apart from the destruction of PAHs, what additional factors might contribute to the apparent reduction of PAH strength in AGN environments?  The equivalent widths of the PAH features can be reduced if the mid-IR spectrum is significantly diluted by the featureless continuum from hot dust emission from the AGN torus. Although this possibility has been raised frequently (\eg Aitken \& Roche 1985; Roche \etal 1991; Genzel \etal 1998; Sturm \etal 2000; Sales \etal 2010; LaMassa \etal 2012; Shi \etal 2014; Xie \etal 2017), to date it has yet to be investigated thoroughly.  Of course, PAH emission may be weak simply because of the dearth of interstellar medium, or because star formation is inefficient nothwithstanding the availability of gas.  The goal of this study is to break these degeneracies using a sample of bright, nearby quasars whose host galaxy interstellar medium has been well characterized (Shangguan et al. 2018, 2020a, 2020b; Molina et al. 2021) and whose star formation properties can be independently ascertained (Xie et al. 2021) with a SFR indicator suitable for AGNs (Zhuang et al. 2019). This work is based on a detailed re-analysis of the mid-IR spectra originally published by Shi et al. (2014).

Section~\ref{sec:sample_data} presents the sample, data, and mid-IR spectral measurements. We investigate the continuum dilution effect in Section~\ref{sec:dilution} and compare different SFRs in Section~\ref{sec:result}. Section~\ref{sec:discussion} discusses the implications of our results, with a summary of the main conclusions given in Section~\ref{sec:conclusion}. This work adopts the following cosmological parameters: $\Omega_m = 0.308$, $\Omega_\Lambda = 0.692$, and $H_{0}=67.8$ km~s$^{-1}$~Mpc$^{-1}$ (Ade \etal 2016). All SFRs reference the stellar initial mass function of Salpeter (1955).

\section{Sample, Data, and Measurements \label{sec:sample_data}}
 
We study the 87 low-redshift ($z < 0.5$) optical/UV-selected quasars from the Palomar-Green (PG) survey (Schmidt \& Green 1983), as summarized in Boroson \& Green (1992).  Being bright and nearby, this sample has been studied extensively in previous works and has accumulated a vast amount of ancillary observations and physical parameters.  As summarized in Shangguan \etal (2018), these include black hole (BH) mass ($M_{\rm BH}$), bolometric luminosity ($L_{\rm bol}$), Eddington ratio ($\lambda_{\rm E}  = L_{\rm bol}/L_{\rm Edd}$, with $L_{\rm Edd}$ the Eddington luminosity), and cold gas content inferred from robust total dust masses.  Stellar masses ($M_*$) are available for the majority of the objects from high-resolution optical and near-IR photometry (Zhang \etal 2016; Zhao et al. 2021), and for the rest they can be estimated from the empirical $M_{\rm BH}-M_*$ relation, which for relatively massive, early-type galaxies has an intrinsic scatter of 0.65~dex (Greene \etal 2020).  

The primary objective of this study is to obtain accurate measurements of the PAH emission for the PG quasars, placing special emphasis on understanding the nature of the sources for which the PAH features are either weak or undetected.  The low-redshift PG quasars have complete low-resolution mid-IR ($5 - 40\, \mum$) spectra acquired by Shi et al. (2014) using the Spitzer Infrared Spectrograph (IRS; Werner \etal 2004)\footnote{PG\,0003+199 only has a low-resolution spectrum covering $5 - 14\, \mum$.  We adopt a high-resolution spectrum available over $14 - 38\, \mum$ and degrade it to lower resolution to measure the PAH emission (Shangguan \etal 2018).}.  The low-resolution mode of IRS has both short and long slits to span a different wavelength range (Houck \etal  2004). The slit size of the short-low mode is $3{\farcs}6\times 57{\arcsec}$ and $3{\farcs}7 \times 57{\arcsec}$, covering, respectively, $5.2-7.7\ \mum$ and $7.4-14.5\ \mum$. The long-low mode covers $14.0-21.3\ \mum$ with aperture size $10{\farcs}5 \times 168{\arcsec}$ and $19.5-38.0\ \mum$ with aperture size $10{\farcs}7 \times 168{\arcsec}$. In each segment, the spectral resolution varies from $\lambda/\Delta\lambda \approx 64$ to 128.  We adopt the aperture-corrected IRS spectra from Shi \etal (2014), who scaled the short-low spectra  to match the long-low data. All the spectra are then scaled to match the MIPS 24~$\mum$ photometry to attain a global measurement. The resultant flux density of the final spectrum is consistent with the WISE (Wright et al. 2010) W3 and W4 bands (see Appendix~A in Shangguan \etal 2018).  We exclude the radio-loud quasar PG~1226$+$023 (3C~273), whose synchrotron-emitting jet can bias the measurements significantly (Shangguan \etal 2018). Our final sample therefore contains 86 quasars with low-resolution spectra (Table~\ref{tab:all}).

We measure the PAH strength using the low-resolution spectra and the methodology of Xie \etal (2018a), which we briefly summarize here.  After masking the ionic emission lines that are not blended seriously with the main PAH features, we fit the $\sim 5-40\ \mum$ spectra with a four-component model consisting of a theoretical PAH template plus a dust continuum represented by three modified blackbodies of different temperatures.  None of the PG quasars suffers from noticeable mid-IR extinction. Instead, most objects show prominent silicate emission around 9.7 and 18~$\mum$, whose peak position shifts toward longer wavelengths (see also Shi et al. 2014), which are modeled with a warm silicate component (Section~4.4 in Xie \etal 2018a). Assuming optically thin conditions in the mid-IR, the silicate emission can be represented by the product of the mass absorption coefficient ($\kappa_{\rm{\nu}}$) and the Planck function $B_{\rm{\nu}}$($T$), where $T$ is a free parameter in the fit. The exact curvature of $\kappa_{\rm\nu}$ depends on the grain size, shape, and chemical composition of silicate dust. Considering grain sizes from 0.1 to 10~$\mum$ and six different chemical species, Xie \etal (2017) find that spherical ``astronomical silicate'' grains of size $\sim$1.5~$\mum$ can describe the width and peak position of the silicate emission in most PG quasars; a few favor micron-sized pyroxene and amorphous olivine. We adopt the best-fit grain size and chemical composition parameters for silicate dust and calculate $\kappa_{\rm{\nu}}$ following Xie \etal (2017). Section~5 further elaborates on the uncertainty of modeling silicate emission. The final model has nine free parameters. We constrain the global best fit with the Levenberg-Marquardt $\chi^2$-minimization algorithm {\tt MPFIT} (Markwardt 2009).  We subtract the best-fit continuum and silicate emission from the observed IRS spectrum to derive the PAH spectrum. We use the integrated PAH strength within $5-15\ \mum$ to calculate the SFR, since this spectral range includes all PAH features and associated continuum within the short-low spectrum and is less sensitive to the ionization and size distribution of the carriers of the individual PAH bands.  For completeness, we also measure several key individual PAH features (6.2, 7.7, 8.6, 11.3, and 17.0~$\mum$) using the Drude profile and parameters given in Table~1 of Draine \& Li (2007).  To estimate the uncertainty of the PAH strength, we use the bootstrap method to sample the spectra multiple times and repeat the fit. The median value of 100 realizations and their corresponding standard deviation give the final PAH flux and its $1\,\sigma$ error. The measurement is regarded as an upper limit if the PAH flux is less than $3\,\sigma$. As all the host galaxies have $M_* \,>\, 10^{9}\,M_{\odot}$ (Table~\ref{tab:all}), we adopt the PAH-based SFR calibration of Xie \& Ho (2019) for massive galaxies, which has a scatter of 0.2\,dex.  Expressing the integrated $5-15\ \mum$ PAH luminosity ($L_{\rm PAH}$) in units of \lum, 

\begin{equation}
\log\ ({\rm SFR_{PAH}}/M_{\odot}\ {\rm yr}^{-1})\,=\,(0.948 \pm 0.034)\, (\log L_{\rm PAH} - 43.0) \ + \,(0.675 \pm 0.024).
\label{eq:sfr_pah}
\end{equation} 

\
\
\

To evaluate the reliability of using PAH emission to estimate SFRs in the context of luminous AGNs, it is crucial that we have access to independent SFR indicators already known to be trustworthy.  To this end, we utilize a subset of 38 PG quasars that also has useful high-resolution IRS observations, which permit us to derive SFRs based on the fine-structure lines of [Ne~II]~12.81~$\mum$ and [Ne~III]~15.56~$\mum$ according to the method of Ho \& Keto (2007), as extended to AGNs by Zhuang et al. (2019).  The measurement of the neon lines and their conversion to \sfrneon\ for the PG quasars are described in Xie et al. (2021). The high-resolution spectra taken with the short-high mode ($9.9 - 19.6\,\mum$) were acquired with an aperture of width $4{\farcs}7$, which, while narrow, fortunately is comparable to that of the short-low spectra.  This enables us to adopt the aperture flux correction factors derived for the low-resolution spectra from Shi \etal (2014). The majority (90\%) of the sources have flux correction factors $\lesssim 20\%$, comparable to the measurement error of the neon flux. We further extend the dynamic range of the SFR comparison using the torus-subtracted total ($8-1000\ \mum$) IR luminosity, which is available for the entire PG quasars.\footnote{Shi \etal (2014) calculated the SFR from the 160~$\mum$ luminosity for the PG sample. Compared with that work, the SED coverage of our study has been expanded significantly from $\sim 5-160\ \mum$ to $\sim 1-500\ \mum$. Moreover, Shangguan \etal (2018) decompose the broadband SED into separate components to account for the starlight, the dusty torus, and the large-scale cool interstellar medium of the host (see Section~3).} Using the total IR luminosity for the host galaxy, Xie \etal (2021) demonstrate that \sfrir\ is consistent with \sfrneon\ within a scatter of $\sim$0.4 dex for quasars with bolometric luminosities $L_{\rm bol} \approx 10^{44.5} - 10^{47.5}\, \rm erg\,s^{-1}$.

\section{Dilution of PAH Features by the AGN \label{sec:dilution}}

\begin{figure}
\begin{center}
\includegraphics[height=0.6\textwidth]{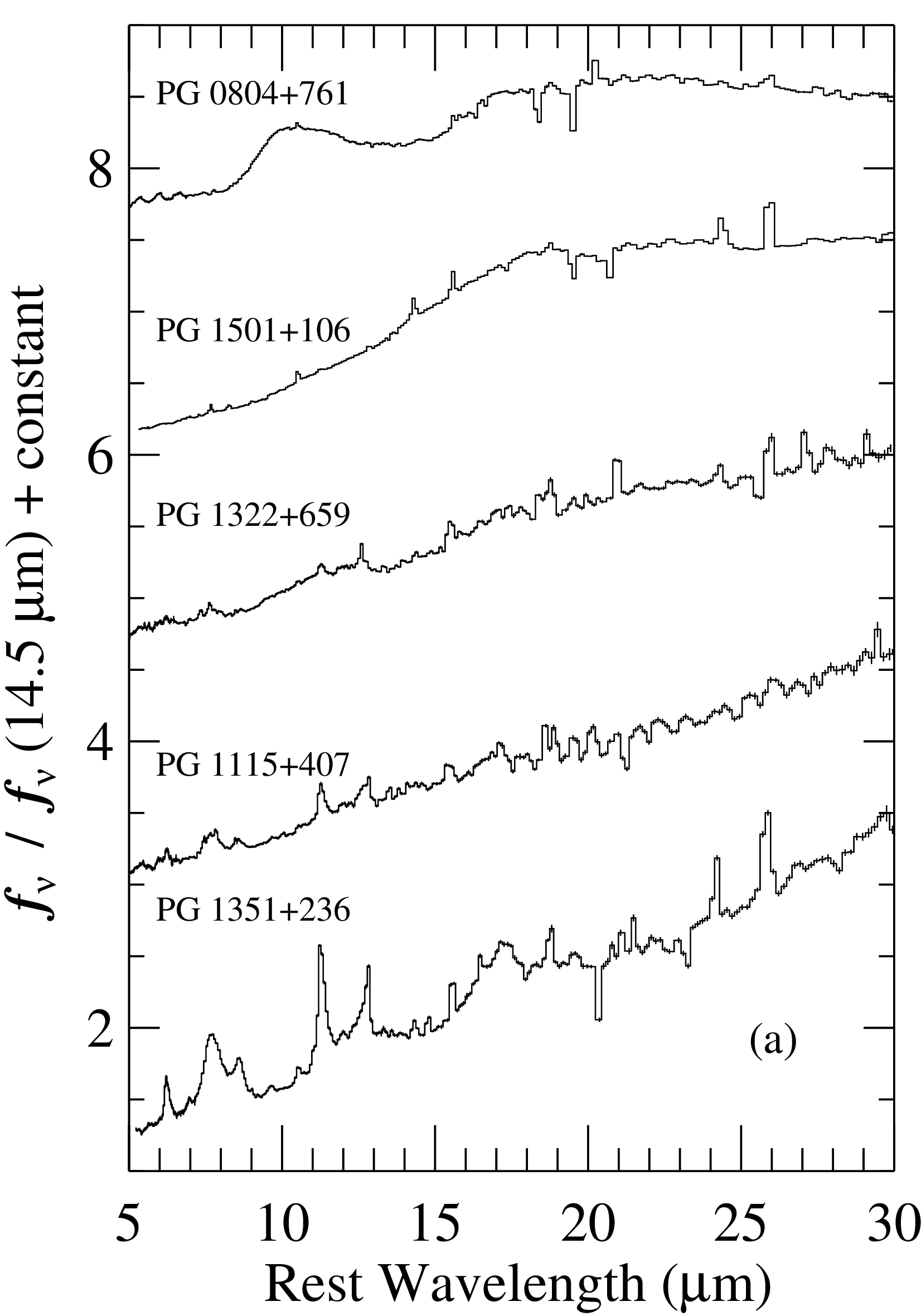} 
\includegraphics[height=0.6\textwidth]{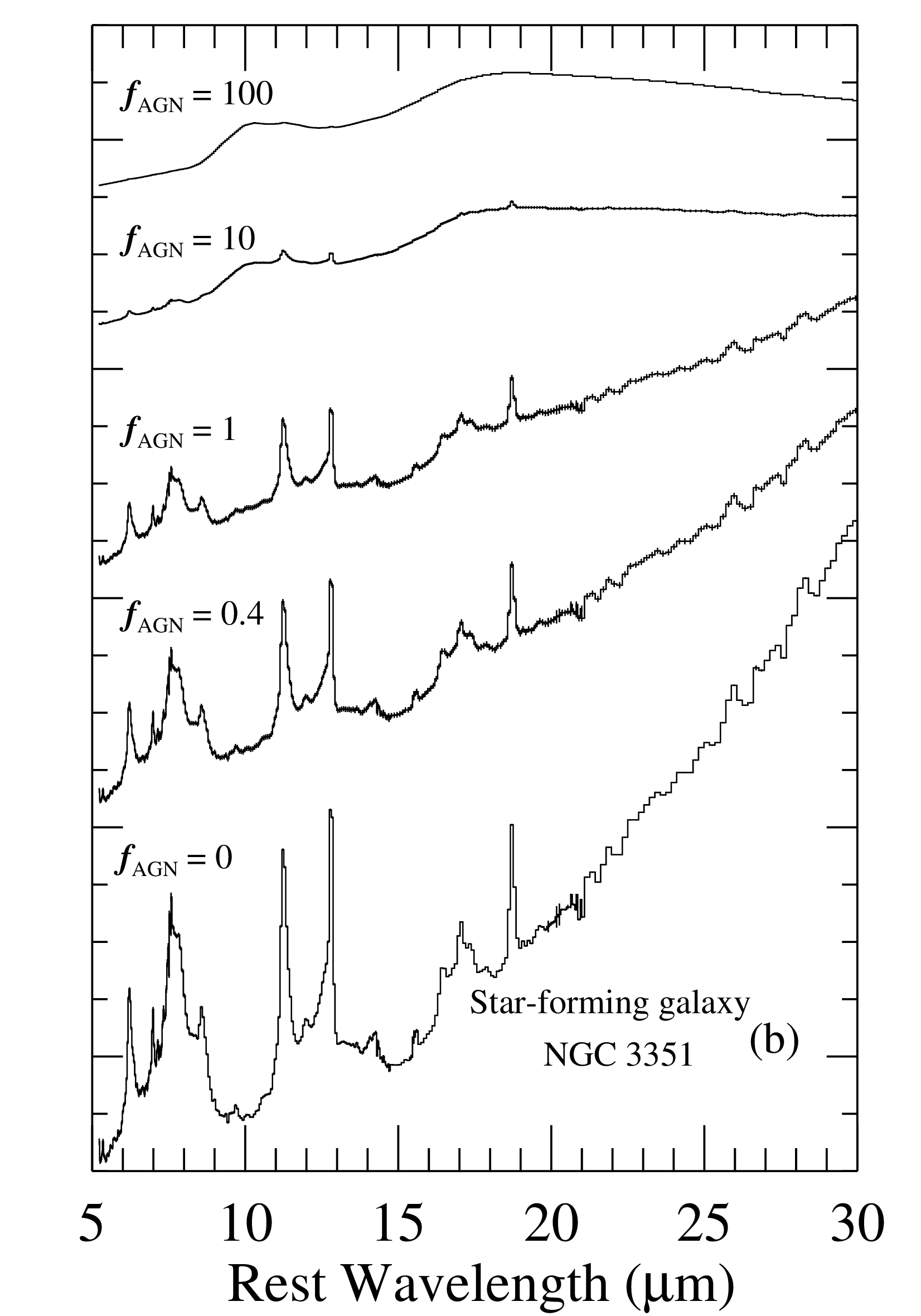}
\caption{The diversity of PAH emission strength (a) observed in the IRS spectra of PG quasars and (b) reproduced from mock spectra in which different AGN fractions ($\agnfrac$) have been added to the pure ($\agnfrac = 0$) SFG NGC 3351. Each spectrum has been normalized to the flux density at 14.5~$\mum$ and offset by an arbitrary constant for illustration purposes.}
\label{fig:pah_agn}
\end{center}
\end{figure}
\noindent

Quasars exhibit a wide diversity of mid-IR spectra, ranging from those with strong PAH emission that closely resemble SFGs, such as PG\,1351$+$236, to continuum-dominated objects that are essentially devoid of line emission, such as PG\,0804$+$761. Figure~\ref{fig:pah_agn}a shows five examples, ordered according to PAH strength.  We conduct a simple experiment to explore the possibility that the reduction of PAH equivalent width in AGNs is caused by the continuum dilution from the dusty torus (\eg Sturm \etal 2000).  This allows us to ascertain quantitatively the effect of continuum dilution and to interpret properly the upper limits of objects with weak or undetected PAH emission.  We simulate spectra covering $5-40\ \mum$ to mimic the IRS observations.  To generate the mock spectra of active galaxies, we first assume that the signal consists of three components: (1) dust emission from the AGN torus; (2) interstellar dust emission from the host galaxy, and (3) stellar emission from host galaxy starlight. We assume that the three components are unresolved and do not consider their spatial distribution.  This assumption is valid because the vast majority of the host galaxies of the PG quasars are unresolved within the IRS slit. Last but not least, we suppose that the three components are distinct from each other and that we do not need to consider radiation transfer. The composite spectrum is then constructed by linear combinations of the three components. In practice, we do not consider host galaxy starlight because in the present sample it contributes $\lesssim 1\%$ to the total mid-IR power on a global scale, indicating that evolved stars do not contribute significantly in the spectral range of interest. Indeed, the torus emission outshines the starlight at $\lambda \gtsim 3~\mum$ (Shangguan \etal 2018; Zhuang \etal 2018). Moreover, our mock simulations adopt the empirical template of an SFG (NGC~3351; see below) to represent the host galaxy, whose spectrum already includes the proper proportion of starlight. Our assumption would not apply to the lower luminosity AGNs, such as those that reside in elliptical galaxies whose old stellar population radiates significantly at $5-20\,\mum$ (\eg Smith \etal 2007; Rampazzo \etal 2013), but this is not the regime that concerns us here. We neglect the line emission contribution from the gas, either from the narrow-line region of the AGN or from the host galaxy, since the isolated narrow lines are removed prior to measuring the PAH emission (Section~\ref{sec:sample_data}).  The lines that blend with PAH ([Ne~II]~12.81~$\mum$ and $\mathrm{H_{2}}\, S$(1)~17.03~$\mum$) only contribute $\sim$2\% to the total PAH luminosity, which is less than the measurement error for the latter ($\sim$20\%).

\begin{figure}
\begin{center}
\includegraphics[width=0.46\textwidth]{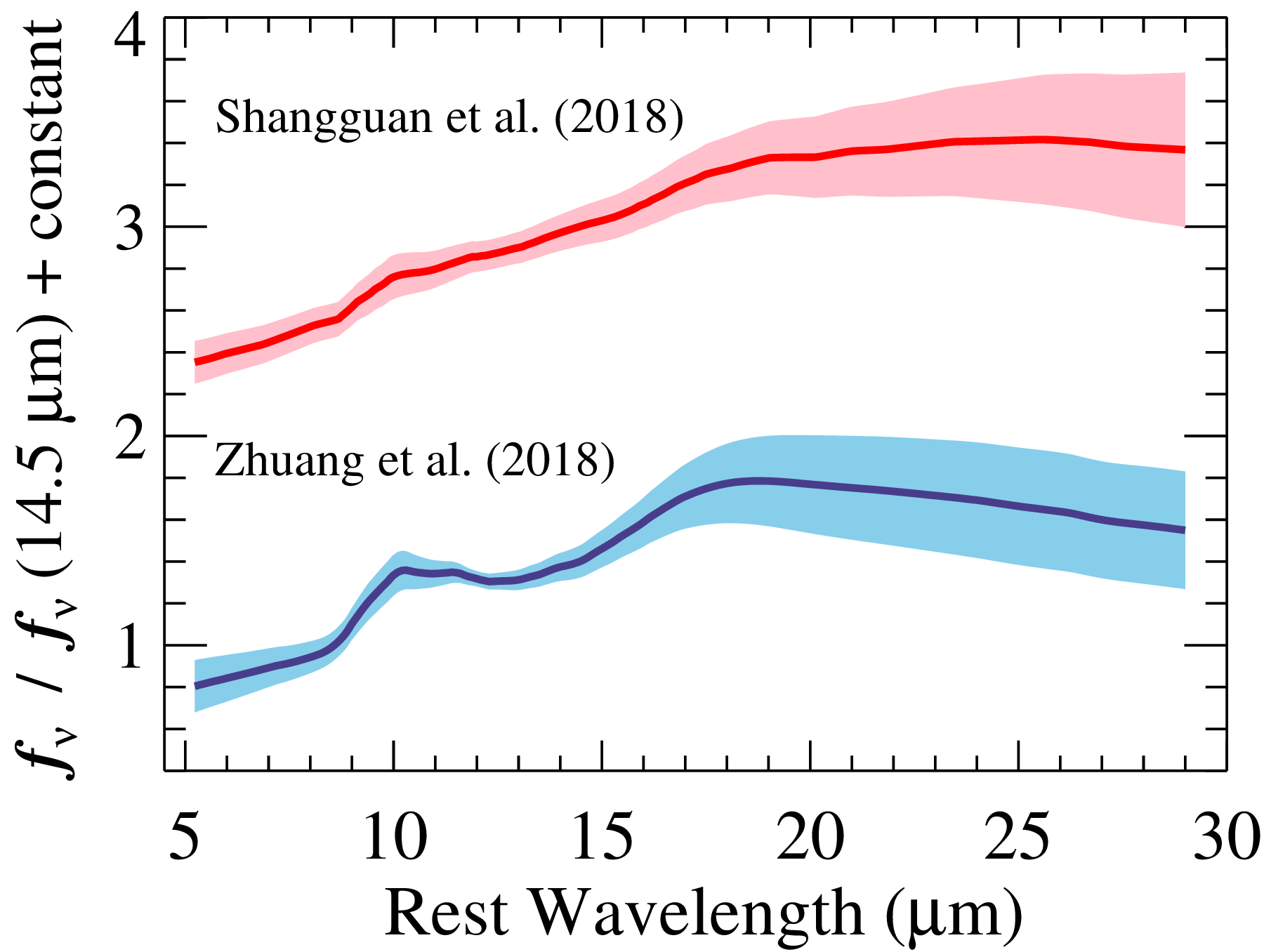}
\caption{The mid-IR spectra of the median AGN torus template from Shangguan \etal (2018; red) and Zhuang \etal (2018; blue).  The shaded region indicates the standard deviation of each template.}
\label{fig:torus_sfg}
\end{center}
\end{figure}

For the AGN continuum, we adopt the median spectrum of the best-fit torus model for the full sample of PG quasars from Shangguan \etal (2018), which is based on the detailed decomposition of the $1-500\, \mum$ SED into a stellar component, a galaxy-scale cool dust component, and a torus component consisting of warm dust calculated from the {\tt CLUMPY} model of Nenkova \etal (2008a, 2008b), supplemented by an additional blackbody component with $T \approx 1000$\,K to account for the hot dust radiation in the near-IR (see Section 4.1 of Shangguan et al. 2018).  We also consider the {\tt CAT3D} wind torus model of H{\"o}nig \& Kishimoto (2017), which Zhuang et al. (2018) used to analyze the PG sample following the methodology of Shangguan \etal (2018).  Compared to {\tt CLUMPY}, {\tt CAT3D} does a better job at fitting the silicate emission profile at 9.7 and 18~$\mum$, and it can account for the near-IR emission without having to introduce an ad~hoc hot dust component.  The two torus models are illustrated in Figure~\ref{fig:torus_sfg}.  We use the observed mid-IR spectrum of NGC~3351 (Figure~1b, bottom), a typical SFG from the SIRTF Nearby Galaxy Survey (SINGS; Kennicutt \etal 2003), to represent the interstellar dust emission of the host galaxy. The average signal-to-noise ratio\footnote{For flux density $f_i(\lambda)$ at wavelength $\lambda$ with uncertainty $\sigma_i(\lambda)$, S/N\,=\,$\sum f_i(\lambda)/ \left(\sum \sigma_i^2(\lambda)\right)^{1/2}$.} of the IRS spectrum of NGC~3351 is $\sim 100$.  Xie \etal (2018a; their Figure~14) find that the PAH spectrum varies little among massive SFGs. The dispersion of the average PAH spectrum of SFGs is less than the PAH measurement uncertainty ($\sim$ 20\%). Thus, adopting a different SFG template will not affect our main conclusions.

The final mock spectra are constructed by summing the host galaxy component [$f_{\rm host} (\lambda)$] with various fractions ($\agnfrac$) of the torus component [$f_{\rm torus} (\lambda)$], with $\agnfrac \equiv \{\int^{40\,\mum}_{5\,\mum} f_{\rm torus}(\lambda)\,{\rm d}\lambda\}/ \{\int^{40\,\mum}_{5\,\mum} f_{\rm host} (\lambda)\,{\rm d}\lambda$\}, after normalizing the torus component to have the same integrated flux as the host at $5-40~\mum$. Defined based on the integrated flux, $\agnfrac$ is largely insensitive to possible variations in individual PAH band ratios among SFGs. We simulate 40 mock spectra with discrete values of $\agnfrac = 0.1$ to 100, which encompass the observed range for the PG sample. The error budget of the mock spectrum consists of 
 
\begin{equation}
\label{eq:sigma_mock}
\sigma^2 (\lambda) = \sigma_{\mathrm{sys}}^2 (\lambda) + 
                     \sigma_{\mathrm{host}}^2 (\lambda) + 
                     \sigma_{\mathrm{AGN}}^2 (\lambda),
\end{equation}
 
\noindent
where $\sigma_{\rm sys} (\lambda)$ represents the systematic errors from instrumental effects (\eg readout noise, bias subtraction, and dark current) and data reduction, and $\sigma_{\mathrm{host}} (\lambda)$ and $\sigma_{\mathrm{AGN}} (\lambda)$ account for the photon noise associated with the host galaxy and torus component, respectively. The first two terms in Equation~\ref{eq:sigma_mock} are included already in the observed spectrum of NGC~3351,  while the last term can be estimated from the quadrature sum of the Poisson noise from the torus emission and the variance of the torus template (delineated by the red and blue shaded regions in Figure~\ref{fig:torus_sfg}):

\begin{equation}
\label{eq:sigma_agn}
\sigma^2_{\mathrm{AGN}} (\lambda) = { \sigma_\mathrm{Poisson}^2 (\lambda) + \sigma^2_{\mathrm{torus}} (\lambda)}.
\end{equation}

\noindent 
Figure~\ref{fig:pah_agn}b shows five reprentative mock spectra for $\agnfrac= \{0, 0.4, 1, 10,  100\}$.  The spectrum with $\agnfrac=0$ is the observed spectrum of the SFG NGC~3351.

\begin{figure}
\begin{center}
\includegraphics[width=0.96\textwidth]{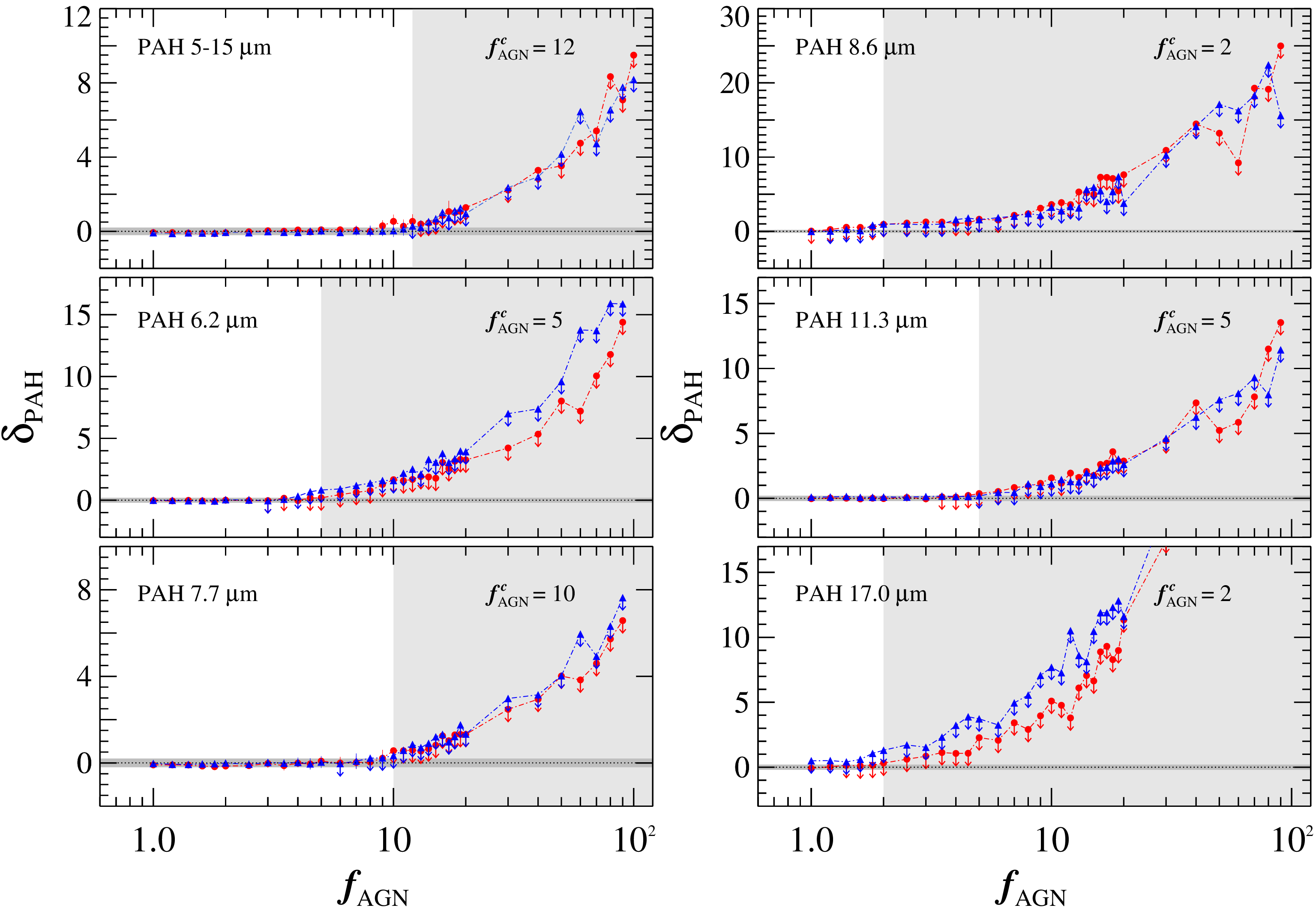}
\caption{Fractional deviation of the recovered PAH strength ($\delta_{\rm PAH}$) as a function of the AGN fraction ($\agnfrac$), for the PAH flux integrated over 5--15~$\mum$ (top left panel) and for individual PAH features (other panels). We show results from the mock spectra using the torus template of Shangguan \etal (2018; red circles) and Zhuang \etal (2018; blue triangles). The gray-shaded zone marks the critical AGN fraction ($\agnfrac^{c}$) above which the torus emission dilutes the PAH measurements by $\delta_{\rm PAH} > 0.2$. }
\label{fig:mock_pah_agnfrac}
\end{center}
\end{figure}

\begin{figure}
\begin{center}
\includegraphics[width=0.96\textwidth]{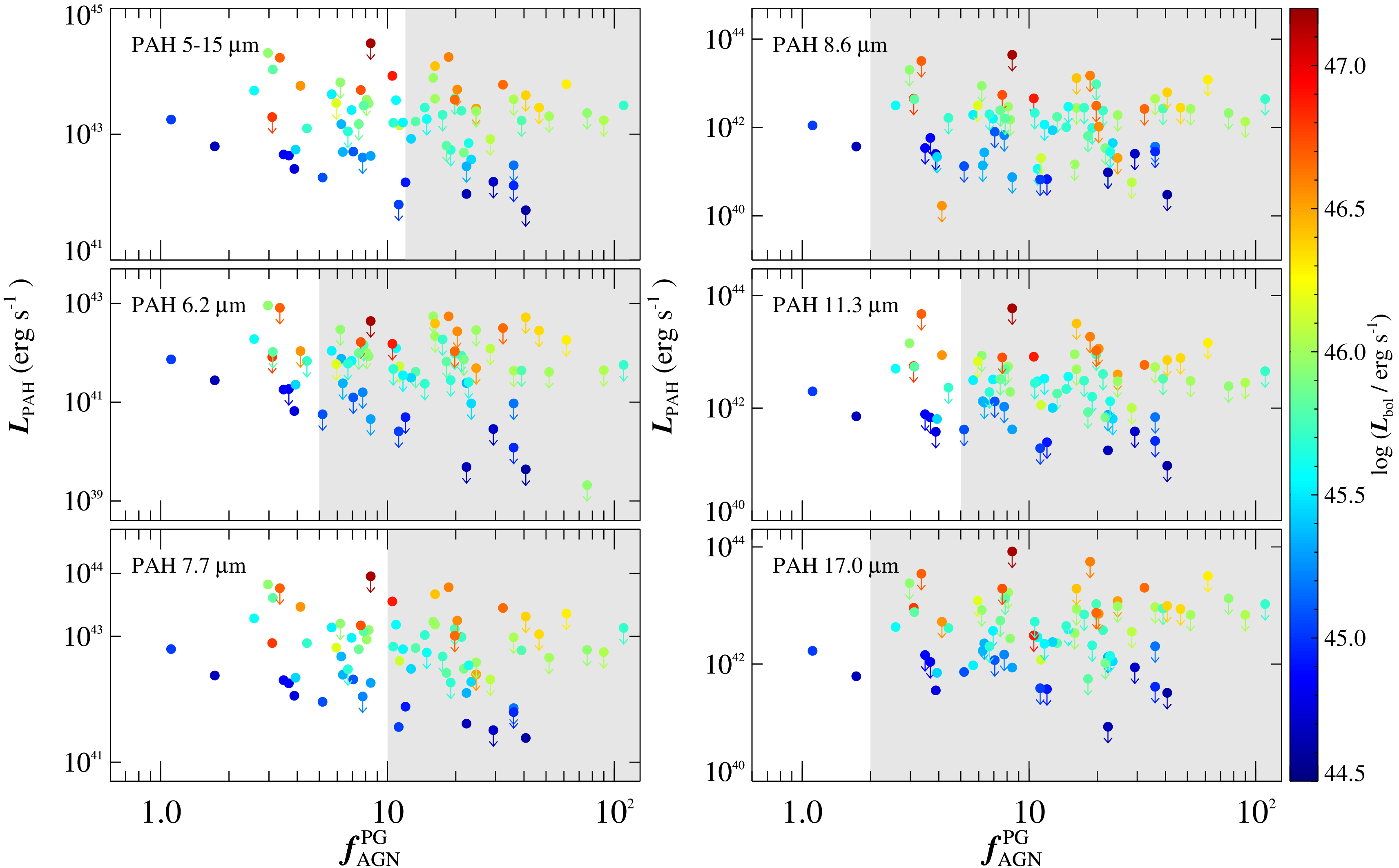}
\caption{Comparison of PAH luminosity with $\agnfracpg$, the AGN fraction directly measured from the SED decomposition of the PG quasars. The data points are color-coded according to \lbol. In each panel, the gray-shaded zone marks the critical AGN fraction ($\agnfrac^{c}$) above which the torus emission dilutes the PAH measurements by $\delta_{\rm PAH} > 0.2$ and therefore are considered unreliable (see Figure 3).  PAH measurements, including upper limits, for objects with $\agnfracpg < \agnfrac^{c}$ are robust.} 
\label{fig:obs_pah_agnfrac}
\end{center}
\end{figure}

We apply the fitting procedure described in Section~\ref{sec:sample_data} to the mock spectra. Given the known PAH flux of the input spectrum of NGC~3351, we measure the output PAH flux of the mock spectra and calculate the quantity 

\begin{equation}
\label{eq:delta_pah}
\delta_{\mathrm{PAH}} = ({\rm PAH}_{\mathrm{output}} - {\rm PAH}_{\mathrm{input}})/{\rm PAH}_{\mathrm{input}}
\end{equation}

\noindent
to evaluate the degree to which the input PAH strength can be recovered. Figure~\ref{fig:mock_pah_agnfrac} illustrates the dependence of $\delta_{\mathrm{PAH}}$ on $f_{\mathrm{AGN}}$ for the two choices of torus templates; the separate panels correspond to the total $5-15~\mum$ PAH flux, as well as the flux of the individual PAH bands at 6.2, 7.7, 8.6, 11.3, and 17.0~$\mum$.  As expected, the mock spectra return robust PAH fluxes ($\delta_{\mathrm{PAH}} \approx 0$) for low values of $f_{\mathrm{AGN}}$, but as $f_{\mathrm{AGN}}$ increases the output PAH fluxes gradually turn to upper limits that are overestimated systematically for large $\agnfrac$. We consider the deviation unacceptable when $|\delta_{\mathrm{PAH}}| \gtrsim 0.2$, which is comparable to the typical PAH flux uncertainty.  The critical AGN fraction above which this threshold is reached is $f_{\rm AGN}^{c} \approx 12$ for PAH~5--15~$\mum$, $f_{\rm AGN}^{c} \approx 5$ for PAH~6.2~$\mum$, $f_{\rm AGN}^{c} \approx 10$ for PAH~7.7~$\mum$, $f_{\rm AGN}^{c} \approx 2$ for PAH~8.6~$\mum$, $f_{\rm AGN}^{c} \approx 5$ for PAH~11.3~$\mum$, and $f_{\rm AGN}^{c} \approx 2$ for PAH 17.0~$\mum$.  PAH upper limits for sources with $f_{\rm AGN} \lesssim f_{\rm AGN}^{c}$ are reliable, but those for $f_{\rm AGN} > f_{\rm AGN}^{c}$ are not because the torus dilution is too severe.

\begin{figure}
\begin{center}
\includegraphics[width=0.48\textwidth]{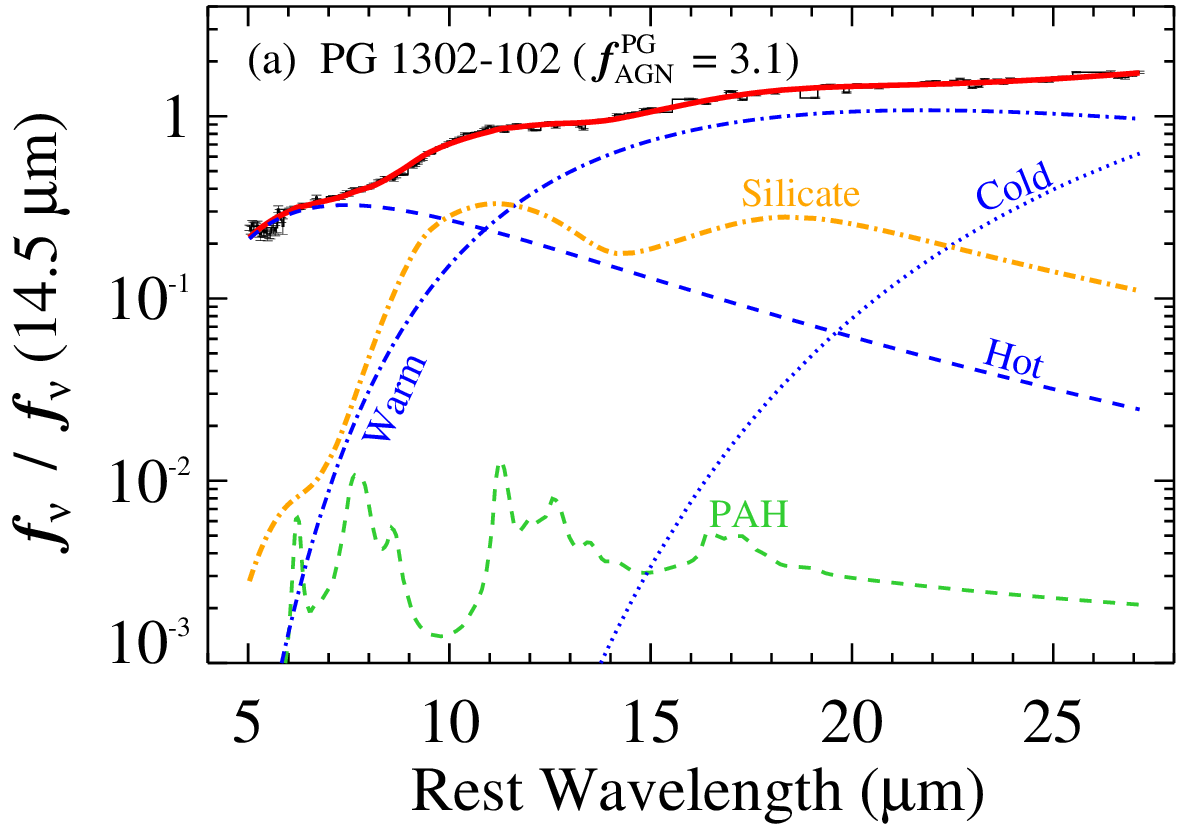}
\includegraphics[width=0.48\textwidth]{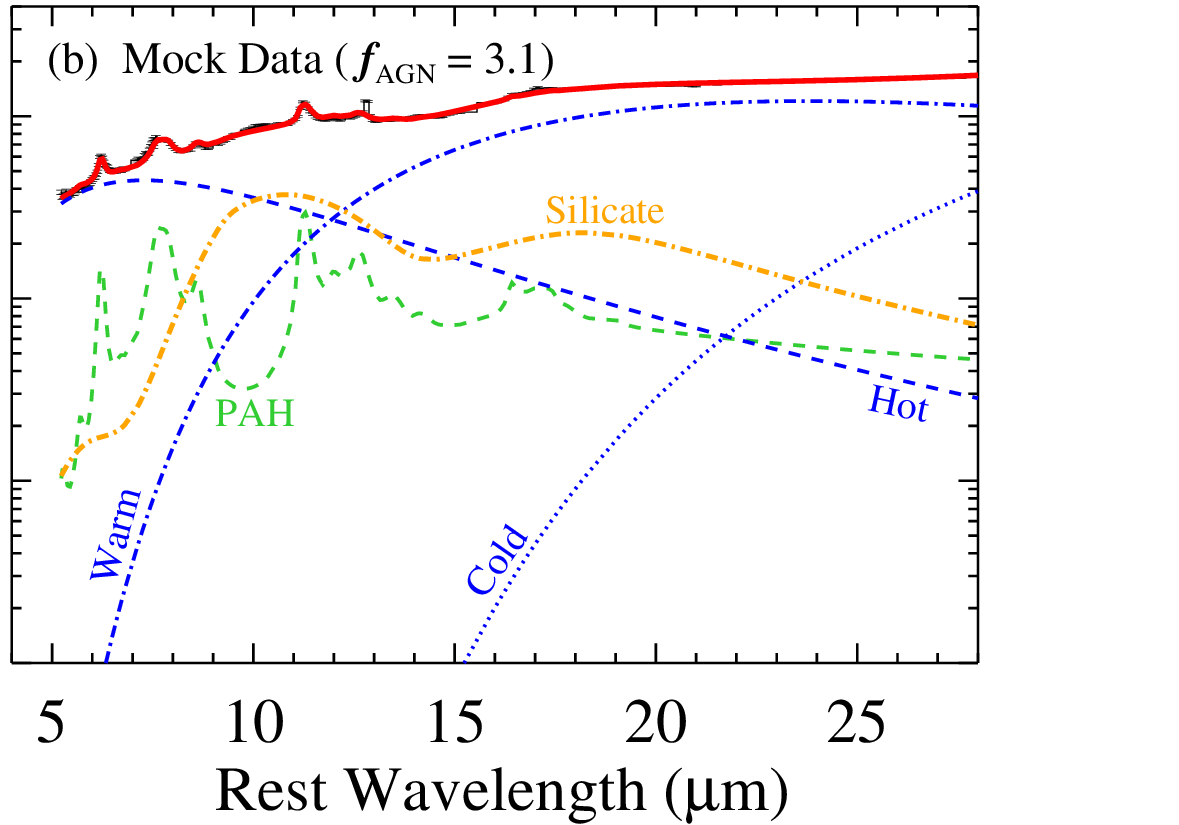}
\caption{Example of PG~1302$-$102, a quasar with intrinsically weak PAH emission and low $\agnfrac$ ($\agnfracpg = 3.1$). Panel (a) shows the observed IRS spectrum (black histogram and error bars), and the best-fit model (solid red line), which consists of three modified blackbodies with hot (blue dashed line), warm (blue dotted-dashed line), and cold (blue dotted line) temperature, the theoretical PAH template (green dashed line), and silicate emission (orange dotted-dashed line).  Panel (b) plots a mock data with $\agnfrac = 3.1$, to mimic the AGN strength observed in PG~1302$-$102. PAH emission is clearly visible in the mock data, but it is absent in the observed spectrum, indicating that PAH emission is intrinsically weak in this quasar.} 
\label{fig:weak_pah_low_fagn}
\end{center}
\end{figure}

Figure~\ref{fig:obs_pah_agnfrac} shows the distribution of PAH measurements as a function of the AGN fraction for the PG quasars, as derived by Shangguan \etal (2018)\footnote{The torus templates of Zhuang \etal (2018) yield consistent results.} from decomposing the 1--500~$\mum$ SED, which consists of photometry from 13 broad bands (2MASS, WISE,  Herschel PACS, and SPIRE) plus low-resolution 5--40~$\mum$ spectroscopic data (Spitzer IRS).  The AGN fraction $\agnfracpg$ is defined as the ratio of the luminosity of the best-fit torus template to the luminosity of the best-fit host galaxy template, integrated over $5-40\,\mum$. Shangguan \etal (2018) modeled the host galaxy component of the PG quasars with the theoretical dust emission template of Draine \& Li (2007), which accurately describes the IR SEDs of SINGS SFGs (including NGC~3351, Draine \etal 2007). Two points of note.  First, PAH emission is detected in a handful of sources with $\agnfracpg > f_{\rm AGN}^{c}$ because the AGN fraction can be highly uncertain in some hosts that have little to no cold dust emission  (e.g., PG~1114$+$445).  Second, the existence of upper limits in sources with $\agnfracpg < f_{\rm AGN}^{c}$ implies that these limits are significant. That is, in these sources PAH emission {\it could}\ have been detected, but it was not. Their PAH strengths are therefore {\it intrinsically}\ weak. We showcase an example in Figure~\ref{fig:weak_pah_low_fagn}.  The observed spectrum of PG~1302$-$102 (Figure~\ref{fig:weak_pah_low_fagn}a) is devoid of obvious PAH features, and a formal decomposition of the spectrum yields a PAH component that is $\sim 2$ orders of magnitude weaker than the continuum.  However, with $\agnfracpg$ = 3.1 (Table~1), according to our mock simulations a quasar with this level of moderate AGN dilution clearly exhibits visible PAH features (Figure~\ref{fig:weak_pah_low_fagn}b) that should have been easily detected, if present.  We stress that our conclusion that PAH can be intrinsically weak or absent in some PG quasars---one of the principal results of this study---critically depends on the assumption that $\agnfracpg$ measured from the data well approximates $\agnfrac$ used in the mock simulations (Figure~\ref{fig:mock_pah_agnfrac}). We believe that this is the case, given that both quantities are defined exactly in the same manner, and the close resemblance between the real and mock spectra (Figure~\ref{fig:pah_agn}).

\begin{figure}
\begin{center}
\includegraphics[width=0.48\textwidth]{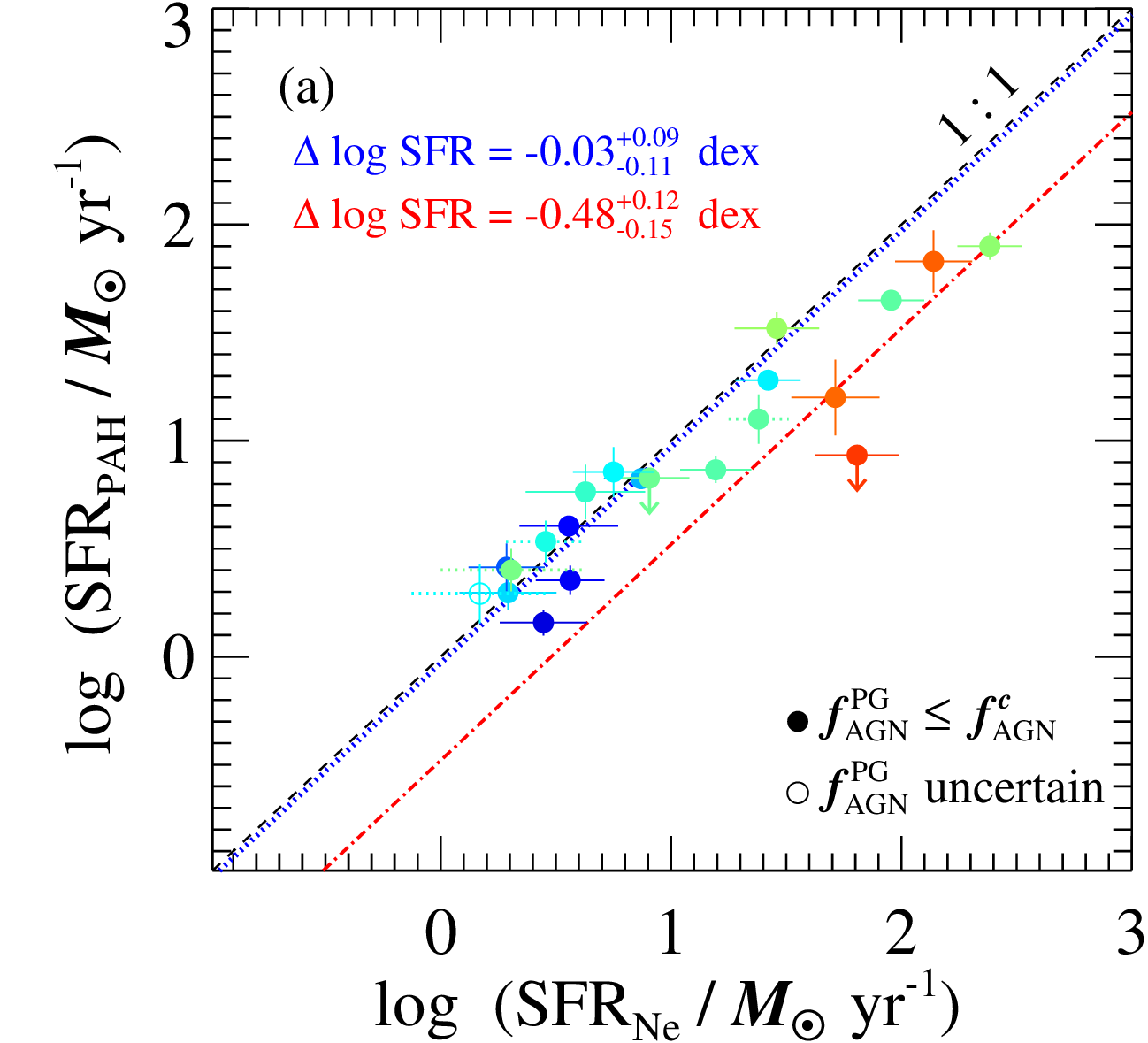}
\includegraphics[width=0.48\textwidth]{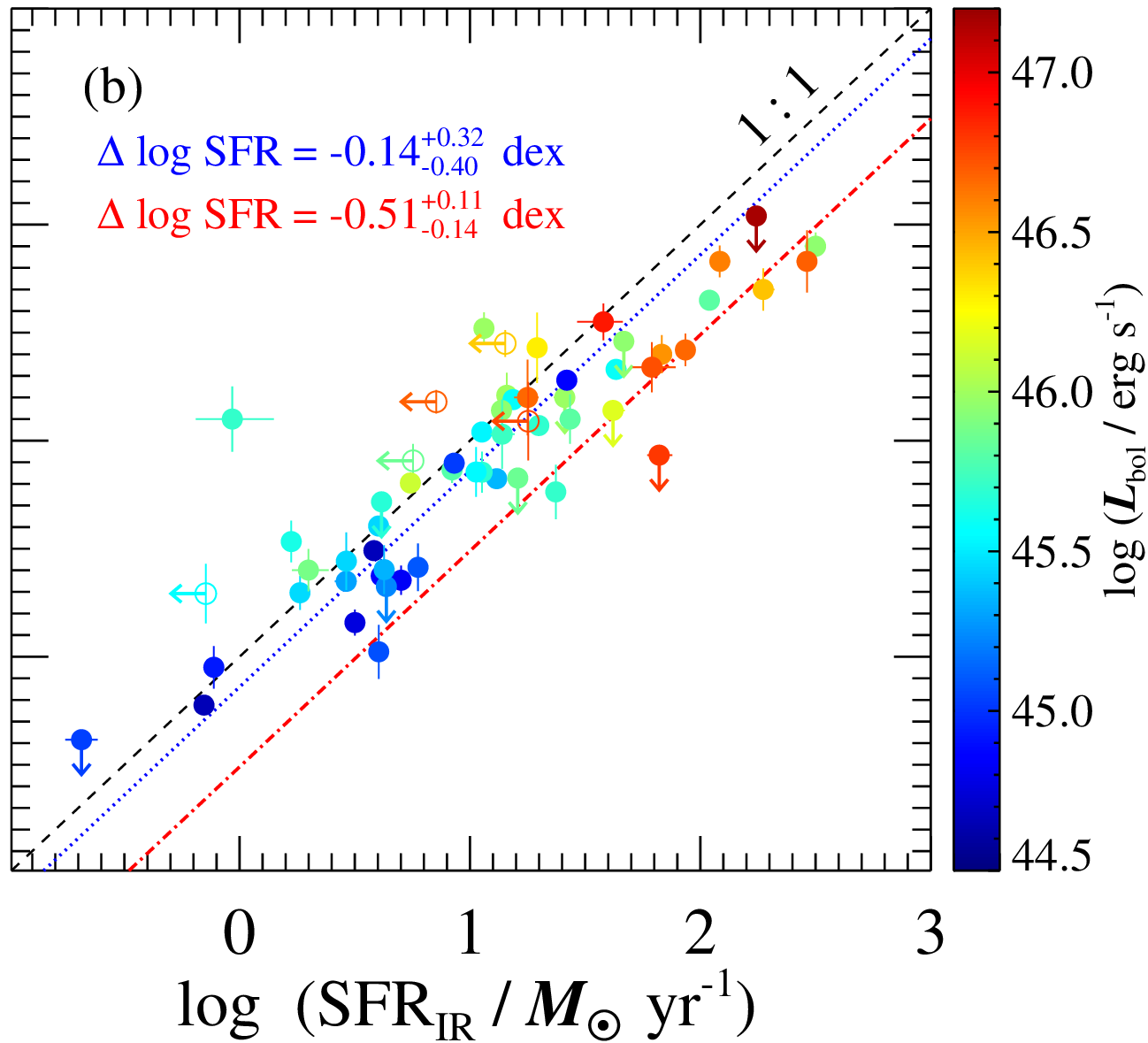}
\caption{Comparison of the SFRs derived using the $5-15~\mum$ PAH luminosity with those obtained from the (a) mid-IR neon lines and (b) total IR luminosity.  Open symbols indicate objects with uncertain $\agnfracpg$ because their IR luminosities are upper limits. The black dashed line represents the one-to-one relation.  The blue dotted line gives the median $\Delta \log {\rm SFR}$ for objects with ${\rm SFR} \lesssim  50\,\usfr$, while the red dotted-dashed line gives the median $\Delta \log {\rm SFR}$ for objects with ${\rm SFR} > 50\,\usfr$.  The points are color-coded according to \lbol, as given in the right color bar. In panel (a), sources for which \sfrneon\ involves upper limits in neon flux are plotted with a dotted line to denote the range of possible values.}
\label{fig:sfr_pah_neon_ir}
\end{center}
\end{figure}

\vskip 30pt

\section{Results \label{sec:result}}

\subsection{The Reliability of PAH-based SFRs \label{subsec:comp_sfr}}

With robust PAH measurements in hand, we can formally compare PAH-based SFRs (Equation~1) with independently estimated SFRs derived from the mid-IR [Ne~II]+[Ne~III] lines, as calibrated by Zhuang \etal (2019).  Figure~\ref{fig:sfr_pah_neon_ir}a shows that \sfrpah\ tracks \sfrneon\ on a one-to-one relation for ${\rm SFR_{Ne}} \lesssim 50\ \usfr$, but that \sfrpah\ systematically underpredicts the true value at higher levels of star formation activity.  Since SFR correlates with \lbol\ (\eg Shangguan \etal 2020a; Zhuang \etal 2021), the PAH deficit in the high-SFR regime also corresponds to quasars of higher luminosity  (\lbol\ $\gtrsim 10^{46}$\,\lum).  Using survival analysis to account for the \sfrpah\ upper limits and the range of \sfrneon\ permitted by the incomplete detection of the neon lines in some objects (see Xie \etal 2021)\footnote{We perform survival analysis using the {\tt Weibullfitter} package in {\tt lifelines.readthedocs.io/en/latest} (Davidson-Pilon et al. 2020).}, the median (16\%, 84\%) of the difference $\Delta \log {\rm SFR} \equiv  \log {\rm SFR_{PAH}} - \log {\rm SFR_{Ne}} = -0.03^{+0.09}_{-0.11}$~dex for ${\rm SFR_{Ne}} \lesssim 50\ \usfr$, consistent with the $\sim 0.2$ dex scatter of the neon (Zhuang \etal 2019) and PAH (Xie \& Ho 2019) SFR calibrations.  For ${\rm SFR_{Ne}} > 50\ \usfr$, $\Delta \log {\rm SFR} = -0.48^{+0.12}_{-0.15}$~dex, although this value is poorly constrained because of the small number of objects in this regime. 
 
Figure~\ref{fig:sfr_pah_neon_ir}b extends the analysis using SFRs obtained using the torus-subtracted, total IR ($8-1000\, \mum$) luminosity.  As with \sfrneon, \sfrpah\ statistically follows \sfrir\ with no significant offset for quasars having low to moderate SFRs, which typically correspond to AGN bolometric luminosities of $L_{\rm bol} \lesssim 10^{46}\, \rm erg\,s^{-1}$, but for more powerful systems the PAH-based technique systematically underestimates the true SFR.  Accounting for upper limits, $\Delta \log {\rm SFR} \equiv  \log {\rm SFR_{PAH}} - \log {\rm SFR_{IR}} = -0.14^{+0.32}_{-0.40}$~dex for ${\rm SFR_{IR}} \lesssim 50\ \usfr$, and $\Delta \log {\rm SFR} = -0.51^{+0.11}_{-0.14}$~dex for ${\rm SFR_{IR}} > 50\ \usfr$. The {\tt logrank\_test}\footnote{lifelines.readthedocs.io/en/latest/lifelines.statistics.html\#lifelines.statistics.StatisticalResult} shows that $\Delta \log {\rm SFR}$ differs significantly between the two regimes, with $p< 0.005$. 

\begin{figure}
\begin{center}
\includegraphics[width=0.9\textwidth]{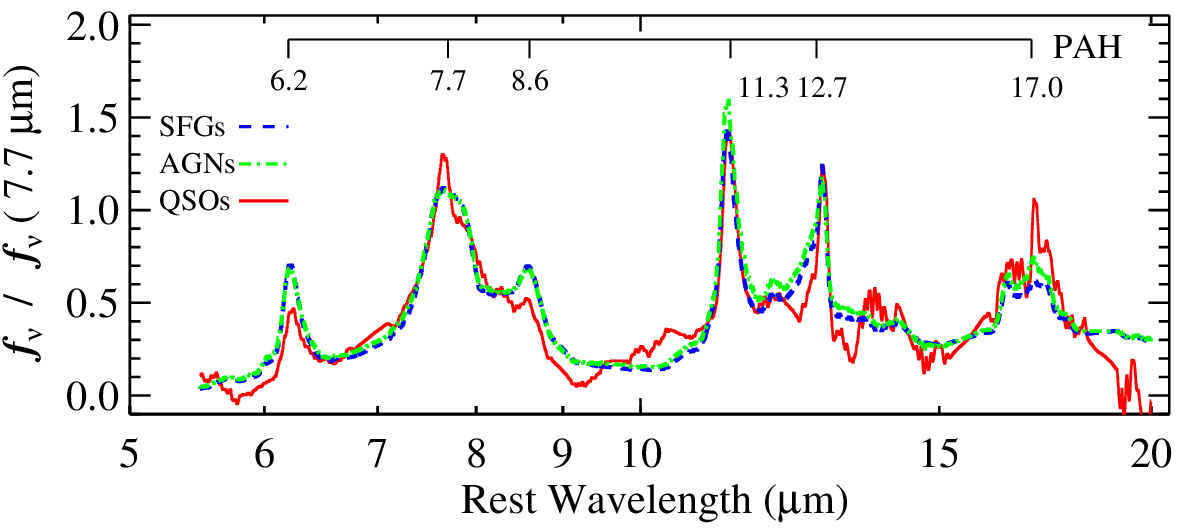}
\caption{Illustration of the mean $5-15~\mum$ PAH spectrum, normalized at 7.7~$\mum$, of PG quasars (red solid line), H~II regions of SFGs (blue dashed line), and the nuclear regions of low-luminosity AGNs (green dash-dotted line). The main PAH emission bands are labeled.}
\label{fig:pahspec}
\end{center}
\end{figure}

\begin{figure}
\begin{center}
\includegraphics[width=0.98\textwidth]{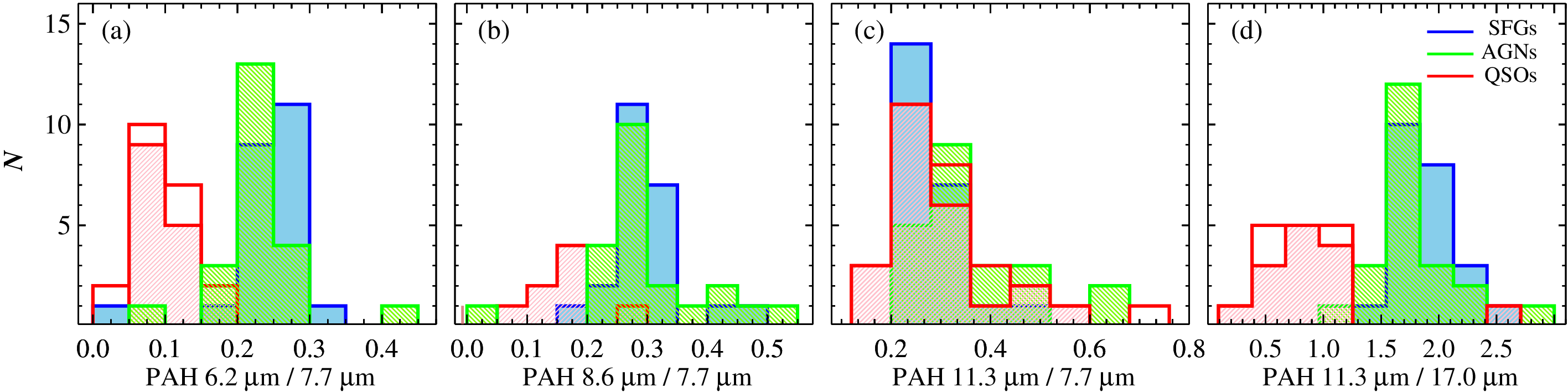}
\caption{Distribution of band ratios for (a) PAH~$6.2\ \mum$/$7.7\ \mum$, (b) PAH~8.6~$\mum$/7.7~$\mum$, (c) PAH~11.3~$\mum$/7.7~$\mum$, and (d) PAH~11.3~$\mum$/17.0~$\mum$ for PG quasars (red), H~II regions of SFGs (blue), and the nuclear regions of low-luminosity AGNs (green). Upper limits that are not affected by torus dilution are included as open histograms.}
\label{fig:hist_pah_ratio}
\end{center}
\end{figure}

\subsection{Mean PAH Spectrum and PAH Band Ratios in Quasar Host Galaxies \label{subsec:pahspec}}

We use the best-fit PAH spectra derived in Section~\ref{sec:sample_data} to construct a mean PAH spectrum for the PG quasar sample (Figure~\ref{fig:pahspec}).  After subtracting the underlying continuum and silicate emission, each individual spectrum is scaled by the mean flux density in the spectral range $7.7 \pm 0.2\ \mum$ prior to computing the error-weighted mean spectrum.  We choose to normalize the spectra at 7.7~$\mum$ in order to highlight potential differences between quasars and other classes of objects in the PAH~$6.2\ \mum$/$7.7\ \mum$ and PAH~$11.3\ \mum$/$7.7\ \mum$ ratios, which are sensitive to PAH size distribution and ionization state (Section~5).  For comparison, Figure~\ref{fig:pahspec} overplots the mean PAH spectrum of the 22 H~II regions from SFGs and the nuclear regions of 23 low-luminosity AGNs acquired from SINGS.  The IRS spectra of the SINGS sources have sufficient signal-to-noise ratio and were analyzed by Xie \etal (2018a) with the same technique as that used for the quasars.  It is immediately apparent that, relative to the comparison sources, quasar host galaxies show markedly weaker 6.2 and 8.6~$\mum$ features, and to a lesser extent the same applies to the 11.3~$\mum$ feature.  This is quantified in Figure~\ref{fig:hist_pah_ratio}, which compares the PAH flux ratio distributions of different galaxy types.  The median and $1\,\sigma$ dispersion of PAH~6.2~$\mum$/7.7~$\mum$ (Figure~\ref{fig:hist_pah_ratio}a) of H~II regions and low-luminosity AGNs are indistinguishable ($0.25\pm0.07$ and $0.24\pm0.07$, respectively). By contrast, quasar hosts have a median (16\%, 84\%) of $0.09^{+0.03}_{-0.03}$.  According to the {\tt logrank\_test}, quasars differ from SFGs and AGNs with $p< 0.005$.  Similarly, quasars exhibit a lower median (16\%, 84\%) PAH~8.6~$\mum$/7.7~$\mum$ ratio of $0.16^{+0.04}_{-0.04}$ relative to H~II regions and low-luminosity AGNs, which have equal values of $0.29\pm0.05$ (Figure~\ref{fig:hist_pah_ratio}b).  The {\tt logrank\_test} yields $p< 0.005$, suggesting different distributions.  Quasars also differ significantly from the comparison groups in terms of PAH~11.3~$\mum$/17.0~$\mum$ (Figure~\ref{fig:hist_pah_ratio}d), having a median ratio of $0.82^{+0.53}_{-0.39}$ that is substantially lower than that of SFGs  ($1.81\pm0.26$) and low-luminosity AGNs ($1.78\pm0.32$), at a significance level of $p< 0.005$.
By contrast, quasars show a median PAH~11.3~$\mum$/7.7~$\mum$ ratio ($0.28^{+0.08}_{-0.07}$) comparable (Figure~\ref{fig:hist_pah_ratio}c; $p= 0.64$) to those of SFGs ($0.27\pm0.06$) and low-luminosity AGNs ($0.3\pm0.12$).

\begin{figure}
\begin{center}
\includegraphics[width=0.98\textwidth]{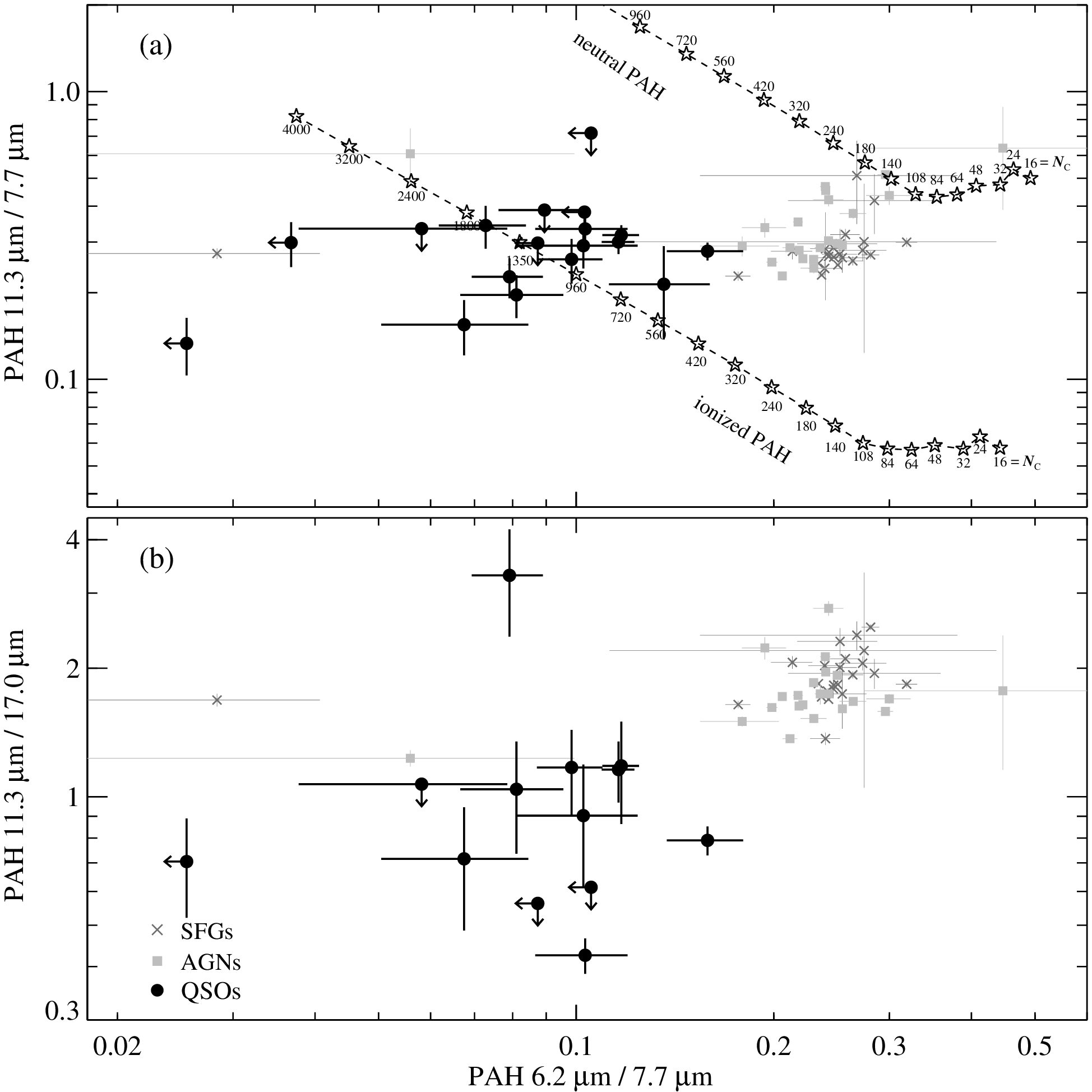} 
\caption{The PAH band ratio diagram of (a) PAH~11.3~$\mum$/7.7~$\mum$ vs. PAH~6.2~$\mum$/7.7~$\mum$ and (b) PAH~11.3~$\mum$/17.0~$\mum$ vs. PAH~6.2~$\mum$/7.7~$\mum$ for quasar host galaxies (circles), H~II regions from SFGs (gray crosses), and the nuclear region of nearby low-luminosity AGNs (filled gray squares). In panel (a), predictions from the theoretical dust models of Draine \& Li (2001) are illustrated as open stars, labeled with the number of carbon atoms ($N_{\rm C}$).  The models assume a Mathis \etal (1983) interstellar radiation field with intensity 1.23 for neutral PAHs and 123 for ionized PAHs. We only include objects with detected PAH~7.7~$\mum$ and 17.0~$\mum$, and none of the 6.2~$\mum$, 11.3~$\mum$ feature is detected when the 7.7~$\mum$ feature is an upper limit. Both detections and non-diluted upper limits are included for the 6.2~$\mum$ and 11.3~$\mum$ features. }
\label{fig:pah_ratio}
\end{center}
\end{figure}

The relative strengths of the PAH bands reflect the size distribution and ionization of interstellar dust grains. Figure~\ref{fig:pah_ratio}a compares the flux ratios of PAH~11.3~$\mum$/7.7~$\mum$ and  PAH~6.2~$\mum$/7.7~$\mum$ with theoretical predictions from Draine \& Li (2001). We adopt their model results for neutral and ionized PAHs of varying grain size exposed to heating intensity strength of 1.23 and 123 times that of the interstellar radiation field of Mathis \etal (1983). We note that the diagnostic plot of  Draine \& Li (2001) is based on PAH strengths that are measured in a manner that closely resembles the PAHFIT methodology of Smith et al. (2007). The method employed for our study, however, yields PAH band measurements that are consistent with those derived from PAHFIT (Xie et al. 2018a), and therefore no systematic bias is introduced. Relative to our comparison sample of SFGs and low-luminosity AGNs, it is immediately apparent that quasars distinguish themselves by their much lower PAH~6.2~$\mum$/7.7~$\mum$ ratios while PAH~11.3~$\mum$/7.7~$\mum$ remains virtually the same, a trend that can be interpreted to mean that quasar hosts have characteristically larger, more ionized grains.  The grain population in quasar hosts contains $N_{\rm C} \approx 900$  carbon atoms or more (some sources formally extend beyond the model grid, which stops at $N_{\rm C} \approx 4000$).  SFGs and weak AGNs can be described by $N_{\rm C} \approx 200$ and lie intermediate between the model sequences for neutral and ionized grains. This interpretation is reinforced further by Figure~\ref{fig:pah_ratio}b.  As with PAH~6.2~$\mum$/7.7~$\mum$, the PAH~11.3~$\mum$/17.0~$\mum$ ratio also traces grain size (\eg Van Kerckhoven \etal 2000), and together they support the notion that quasars are deficient in small grains.

\section{Discussion \label{sec:discussion}}

The degree to which AGNs manifest PAH emission and the main physical reason behind its apparent suppression have long been subjects of investigation.  Early ground-based $8-13~\mum$ spectroscopy of small samples of AGNs noted that their spectra are dominated by a featureless continuum (\eg Kleinmann \etal 1976; Aitken \etal 1981; Cutri \etal 1981; Roche \etal 1984; Aitken \& Roche 1985). In their comprehensive mid-IR atlas of 60 nearby galaxies, Roche \etal (1991) found that objects with active nuclei rarely exhibit narrow emission bands---later recognized as PAHs (Hudgins \& Allamandola 1999)---in contrast with H~II region nuclei, in which these narrow emission bands are pervasive.  Voit (1991) suggested that X-ray heating by AGNs efficiently evaporates the tiny grains ($< 10 \Angstrom$) responsible for the PAH emissions into the gas phase. Unless shielded by dark clouds, PAH molecules of $N_{\rm C} < 100$ in the host galaxy can be completely destroyed by the harsh radiation of the AGN (Voit 1992).  Photo-destruction from X-rays has also been invoked to account for the absence of PAHs in protoplanetary disks (Siebenmorgen \& Kr$\ddot{\rm u}$gel 2010).  With the advent of ISO and especially Spitzer, the vast improvement in the quantity and quality of mid-IR spectra has led to widespread investigations of PAH emission in galaxies and renewed interest in the physical nature of PAHs in AGN environments.  Many studies report that in AGNs PAH emission at $6-10~\mum$ systematically drops relative to emission features at longer wavelengths (\eg Smith \etal 2007; Diamond-Stanic \& Rieke 2010; Sales \etal 2010). The PAH~11.3~$\mum$ feature seems more robust against the influence of the AGN and appears to be a promising SFR indicator. However, debate has persisted over the relative importance of intrinsic destruction of PAHs compared to their mere apparent weakness due to continuum dilution (\eg Genzel \etal 1998; Lutz \etal 1998; Clavel \etal 2000; Sturm \etal 2000; Smith \etal 2007; LaMassa \etal 2012; Alonso-Herrero \etal 2014).  Another explanation for the PAH deficit in AGNs is that active galaxies may simply be gas-poor, as might be expected if energy feedback from the active nucleus effectively expels the interstellar medium.  Or, perhaps they have abundant gas but stars somehow form inefficiently.  

The analysis presented in this study for the PG quasars, in conjunction with recent studies of their gas and star formation properties, sheds considerable light on the nature of PAH emission in AGNs.  The weakness of PAH emission cannot be attributed to the lack of interstellar medium or star formation.  The PG quasars possess a gas content largely commensurate with that of SFGs of similar stellar mass (Shangguan \etal 2018, 2020a), and the gas forms stars with efficiency at least as high as that of galaxies on the star-forming main sequence (Shangguan \etal 2020b; Xie \etal 2021).  In fact, such conditions appear to be generic to other samples of AGNs accreting at moderately high mass accretion rates (e.g., Xia \etal  2014; Xu \etal 2015; Husemann \etal 2017; Kakkad \etal 2017; Stanley \etal 2017; Shangguan \& Ho 2019; Jarvis \etal 2020; Kirkpatrick \etal 2020; Yesuf \& Ho 2020; Zhuang \& Ho 2020).  Instead, we find compelling evidence that quasars genuinely do destroy and modify the grain population responsible for PAH emission (see Section 5.1 and 5.2).

\subsection{PAH Destruction in Quasar Host Galaxies\label{subsec:pahdstr}}

Our analysis of mock spectra (Figure~\ref{fig:pah_agn}b) rigorously quantifies the impact of continuum dilution on the detectability of PAH emission in quasars under conditions that closely mimic those encountered in the host galaxies of the PG sample. Our mock spectra adopt an observationally motivated template for the underlying host galaxy light, as well as the most up-to-date median spectral templates for the AGN torus emission obtained from the PG quasars themselves (Shangguan et al. 2018; Zhuang et al. 2018; Figure~\ref{fig:torus_sfg}).  This then allows us to establish an empirical threshold on the critical AGN-to-host flux ratio ($\agnfrac^{c}$) below which PAH measurements can be considered trustworthy.  The value of $\agnfrac^{c}$ varies from feature to feature (Figure~\ref{fig:mock_pah_agnfrac}), with $\agnfrac^{c} \approx 12$ for the total PAH emission integrated over $5-15\,\mum$, the measure used by Xie \& Ho (2019) to derive SFRs.  

Figure~\ref{fig:obs_pah_agnfrac} reveals a critical finding: although PAH emission could have been detected for all sources with $\agnfracpg < \agnfrac^{c}$, nevertheless there are several sources for which PAH was {\it not}\ detected.  As these PAH upper limits are robust, they are highly significant in guiding our interpretation. There are nine quasars with upper limits in PAH ($5-15\,\mum$). One object (PG~2304+042) has a relatively high AGN fraction ($\agnfracpg$ = 11.2), and therefore its PAH content may be intrinsically low; it happens to be gas-poor ($M_{\rm{gas}} \approx 10^{8.4}\,M_\odot$), and it has very little star formation (\sfrir\ = $0.21\, \usfr$). The other eight sources, however, have $\agnfracpg\ < 10$, $M_{\rm{gas}} = 10^{9.5}-10^{10.6}\,M_\odot$ (Shangguan et al. 2018), and ${\rm SFR_{IR}} = 4-174\,\usfr$, with a median value of $42\,\usfr$. They should have strong, detectable PAH emission, but they do not. Figure~\ref{fig:weak_pah_low_fagn} highlights the particular case of PG~1302$-$102.  Normal SFGs with ${\rm SFR} \gtrsim 1\,\usfr$ already display prominent PAH emission (\eg Smith \etal 2007; Xie \& Ho 2019).  PAH emission should have been detected easily in these quasars, but they are not. We therefore conclude that the PAH carriers must have been destroyed in these systems.   Moreover, the availability of independent SFR estimates for the entire sample permits us to generalize this result (Figure~\ref{fig:sfr_pah_neon_ir}b). In host galaxies with low to moderate SFRs ($\lesssim 50\ \usfr$), which roughly correspond to quasars with $L_{\rm bol} \lesssim 10^{46}\, \rm erg\,s^{-1}$, \sfrpah\ is suppressed relative to \sfrir\ by $\sim$0.14~dex, which may be regarded as statistically insignificant in view of the 0.4~dex uncertainty of \sfrir\ (Xie \etal 2021), but for ${\rm SFR} \gtrsim 50\ \usfr$ or $L_{\rm bol} \gtrsim 10^{46}\, \rm erg\,s^{-1}$, PAH {\it systematically}\ underestimates the true SFR by $\sim 0.5$~dex.  The effect could be even more severe for quasars with luminosities higher than those probed by the relatively local PG sample.  These results are consistent with Shangguan \etal (2018; see their Section~5.2), who, based on modeling the global IR SED, report that PG quasars tend to present lower PAH mass fraction than SFGs, and that the reduction in PAH mass fraction may increase toward more luminous quasars.  

\begin{figure}
\begin{center}
\includegraphics[width=0.9\textwidth]{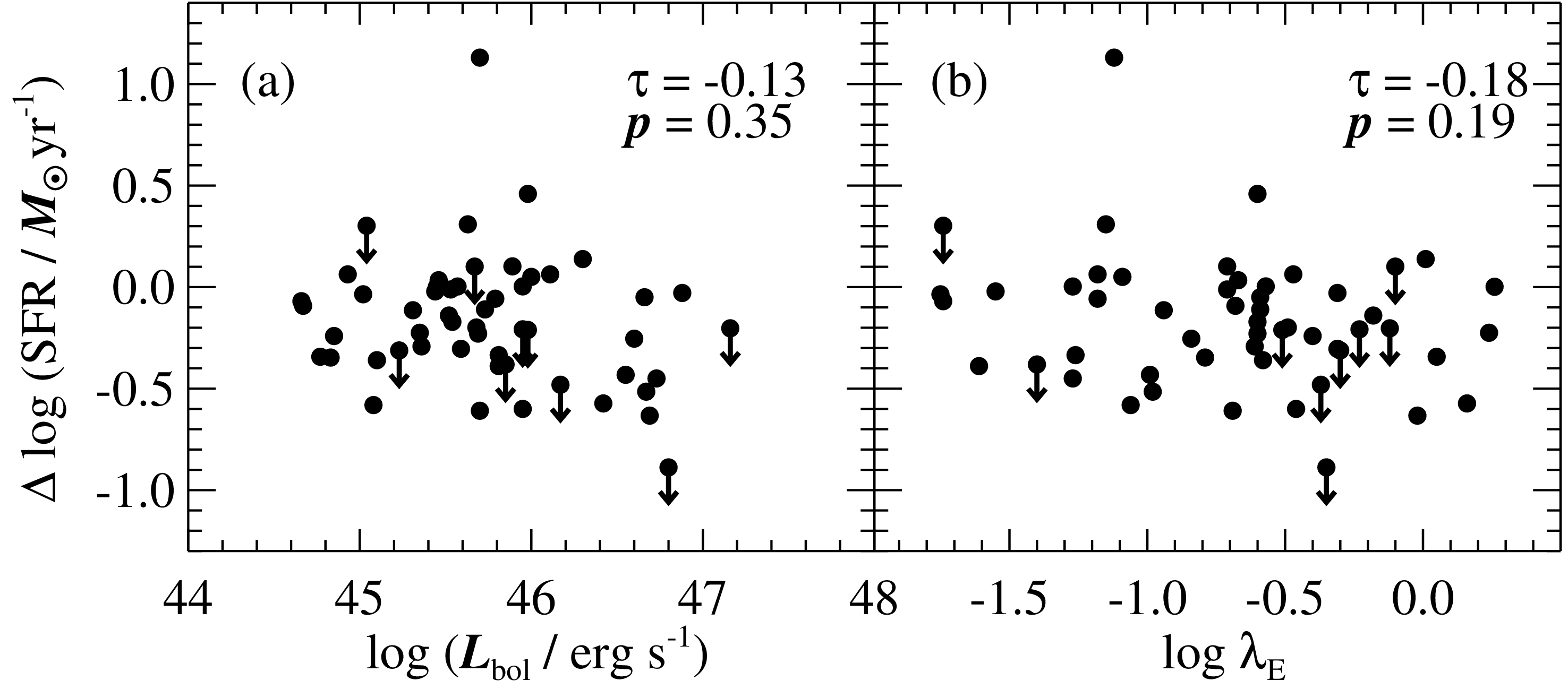}
\caption{Difference between PAH-based and IR-based SFR, $\Delta \log {\rm SFR} \equiv  \log {\rm SFR_{PAH}} - \log {\rm SFR_{IR}}$, versus (a) bolometric luminosity (\lbol) and (b) Eddington ratio ($\lambda_{\mathrm{E}}$). The correlation coefficient $\tau$ and its corresponding $p$ value from {\tt linmix} are given in the top-right corner of each panel. We only include objects with \sfrpah\ detections and upper limits that are not affected by AGN dilution.}        
\label{fig:delta_sfr_agn}
\end{center}
\end{figure}

Does the destruction of PAHs among the most powerful quasars originate from the hard radiation field of the AGN itself, or perhaps from the extreme conditions of violent star formation in the host galaxy?  After all, powerful quasars coexist with powerful starbursts (Xie \etal 2021), and the star formation appears to be centrally concentrated (Zhuang \& Ho 2020; Molina et al. 2021).  We suggest that the AGN plays a more dominant role.  Among massive, starburst-dominated ultraluminous IR galaxies, whose SFRs can reach $1000\,\usfr$ or more, the extinction-corrected PAH-based SFRs faithfully follow the independent SFRs derived from the mid-IR fine-structure neon lines (Xie \& Ho 2019).   In these systems, there is no hint that the intense star formation affects the integrated PAH strength on galactic scales. Thus, we conclude that the hard photons from BH accretion are the main culprit for reducing the PAH emission in quasars.  Figure~\ref{fig:delta_sfr_agn} examines the relation between $\Delta \log {\rm SFR} \equiv  \log {\rm SFR_{PAH}} - \log {\rm SFR_{IR}}$ and the quasar bolometric luminosity (\lbol) and Eddington ratio ($\lambda_\mathrm{E}$).  Although $\Delta \log {\rm SFR}$ shows a very mild inverse trend with \lbol\ and $\lambda_\mathrm{E}$, the correlation is statistically insignificant\footnote{We regard correlations with $p< 0.05$ as statistically significant. The correlation analysis is performed using the {\tt IDL} procedure {\tt linmix} (Kelly 2007), which can consider upper limits for the dependent variable.}.  A larger sample covering a broader dynamic range in quasar properties is needed for future study.

In their comprehensive study of the PG quasar sample, Shi et al. (2014) concluded that SFRs from PAH 11.3~$\mum$ are consistent with those derived from the total IR luminosity, which the authors estimate from matching a star-forming template to the observed photometry at 160~$\mum$. While we agree that SFRs for quasars can be estimated from PAH emission, we find that PAH-based SFRs may be systematically underestimated for ${\rm SFR}\gtrsim 50\,\usfr$. Our analysis differs in two respects compared to that of Shi et al. (2014).  First, we rigorously address the important point of why certain objects have apparently weak PAH emission, clearly distinguishing between those whose PAH features are intrinsically weak from those that are merely diluted by the AGN. And second, instead of focusing on any individual PAH feature, we prefer to use the PAH emission integrated over $5-15~\mum$ because the total emission is less sensitive to band-to-band variations.

\subsection{PAH Ionization and Grain Size Distribution \label{subsec:discus_r_pah} } 

The ionization and size distribution of the grain population determine the relative strength of the different PAH bands. Upon absorption of a single photon, the radiation emitted by small PAH grains peaks at shorter wavelength than that of large grains because smaller grains acquire a higher temperature (\eg Allamandola \etal 1989; Draine \& Li 2001, 2007). Models attribute the 6.2 and 7.7~$\mum$ bands to the C-C stretching mode of ionized PAHs, while the PAH~11.3~$\mum$ band mainly arises from the C-H out-of-plane bending mode of neutral grains (\eg L\'eger \& Puget 1984; Allamandola \etal 1985; DeFrees \etal 1993; Draine \& Li 2001; Hudgins \& Allamandola 2004; Tielens 2008). Given that the  6.2 and 7.7~$\mum$ bands originate from the same vibrational mode of ionized grains, the PAH~6.2~$\mum$/7.7~$\mum$ ratio is sensitive to the grain size distribution. The PAH 11.3~$\mum$/7.7~$\mum$ ratio traces the ionization fraction, in addition to the grain size.

Quasar hosts have systematically lower PAH~6.2~$\mum$/7.7~$\mum$ ratios than SFGs and low-luminosity AGNs.  The median PAH~6.2~$\mum$/7.7~$\mum$ of $0.09\pm0.03$ suggests large PAH molecules of size $N_{\rm C} \gtrsim 900$ (Section~\ref{subsec:pahspec}). A similar, albeit less extreme, conclusion was reached by Sales \etal (2010), who studied low-redshift Seyferts with 12~$\mum$ luminosity $\sim 10^{42}-10^{46}\, \rm erg\,s^{-1}$ (Rush \etal 1993).  The 6.2~$\mum$/7.7~$\mum$ ratios of $0.1-0.2$ found in their sample are consistent with relatively large PAH grains of $N_{\rm C} > 180$.  Our results provide a sharp contrast to previous studies that treated mainly lower-luminosity AGNs.  For example, the very nearby Seyferts selected by Diamond-Stanic \& Rieke (2010) from the Revised Shapley-Ames catalog (Sandage \& Tammann 1987) have PAH ratios of 6.2, 7.7, and 8.6~$\mum$ that are indistinguishable from those of nuclear and off-nuclear H~II regions. O'Dowd \etal (2009) studied low-redshift AGNs and AGN+H~II composites and concluded that their PAH~6.2~$\mum$/7.7~$\mum$ ratios broadly agree with those of SFGs.  The same holds for the work of Wu \etal (2010).  Our present emphasis on powerful quasars best elucidates the destructive effects of AGNs on small PAH molecules (Voit 1992). 

Unlike PAH~6.2~$\mum$/7.7~$\mum$, the distribution of the PAH~11.3~$\mum$/7.7~$\mum$ ratio of quasars closely resembles that of SFGs. By comparison, previous studies that focused on weaker sources generally stressed that PAH~11.3~$\mum$/7.7~$\mum$ is elevated in AGNs\footnote{The previous literature conventionally uses the ratio PAH~7.7~$\mum$/11.3~$\mum$, but in this study we adopt its inverse for ease of comparison with models (e.g., Figure~\ref{fig:pah_ratio}a), and we quote the literature results accordingly.} (\eg Smith \etal 2007; O'Dowd \etal 2009; Diamond-Stanic \& Rieke 2010; Wu \etal 2010; Garc\'ia-Bernete \etal 2021).  Multiple hypotheses have been advanced to account for the higher PAH~11.3~$\mum$/7.7~$\mum$ ratio in active galaxies.  For instance, the overall distribution of grain sizes may be depleted in small grains by the harsh radiation field of the AGN (\eg Smith \etal 2007; O'Dowd \etal 2009; Wu \etal 2010).  Alternatively, the grain distribution may be selectively enhanced in large grains if grain growth is promoted in the rich media associated with the AGN torus (Diamond-Stanic \& Rieke 2010). Another possibility is that the enhanced electron density in the nuclear region may result in a proportionate drop in PAH ionization fraction, leading to a boost of the 11.3~$\mum$ feature relative to the features at shorter wavelength (\eg O'Dowd \etal 2009). Diamond-Stanic \& Rieke (2010) even postulated that AGN-driven shocks can produce more solo C-H groups, which would elevate the 11.3~$\mum$ feature.  On the other hand, upon extending the luminosity range of the sample of O'Dowd \etal (2009) to include more powerful AGNs, LaMassa \etal (2012) reached the opposite conclusion, that AGNs actually present lower PAH~11.3~$\mum$/7.7~$\mum$ ratios and reduced 11.3~$\mum$ equivalent widths.  They argued that AGNs destroy even the neutral PAHs responsible for the 11.3~$\mum$ feature. Apart from sample selection effects, LaMassa \etal (2012) also attributed their different conclusions in part to differences in spatial coverage compared to previous studies\footnote{LaMassa \etal (2012) did not analyze the PAH~6.2~$\mum$/7.7~$\mum$ ratio.}.  Sales \etal (2010) similarly used the 11.3~$\mum$/7.7~$\mum$ ratio to suggest that PAH molecules are more ionized in luminous AGN hosts. 

Our study, which extends the luminosity range even further with the addition of the PG quasars, offers a fresh perspective on this long-standing problem.  For a given grain size specified by the 6.2~$\mum$/7.7~$\mum$ ratio, the 11.3~$\mum$/7.7~$\mum$ ratio of quasars is located far below the model predictions for neutral PAHs. The PAH ratios of quasar hosts occupy the branch of large ($N_{\rm C} \gtrsim 900$), ionized grains (Figure~\ref{fig:pah_ratio}a).  Solely destroying small grains cannot explain the distribution of points for the quasars on this line-ratio diagnostic diagram, as it would increase 11.3~$\mum$/7.7~$\mum$ while reducing 6.2~$\mum$/7.7~$\mum$. Modifying the C-H/C-C ratio would not work either because this process does not affect the 6.2~$\mum$/7.7~$\mum$ ratio (Diamond-Stanic \& Rieke 2010).  We argue that both small grain destruction {\it and}\ ionization fraction enhancement are needed.  The ionization fraction depends on the balance between the photoionization rate and the electron capture rate of ionized PAHs. The exact form scales with $U a \sqrt{T_e}/n_e$, where $U$ is a dimensionless intensity parameter describing the energy density of the radiation field heating the dust, $a$ denotes PAH grain size, $n_e$ is the electron density, and $T_e$ is the electron kinetic temperature (Draine \etal 2021). Draine (2011) shows that for a given size distribution increasing the ionization fraction from 0 to 1 lowers the 11.3~$\mum$/7.7~$\mum$ ratio by a factor of $\sim$6. Similarly, Galliano \etal (2008; their Figure~18b) find that 11.3~$\mum$/7.7~$\mum$ is 8 times lower for galaxies whose ionization fraction is an order of magnitude higher.  At the median value of 6.2~$\mum$/7.7~$\mum$ (or grain size) observed in quasars, the median value of 11.3~$\mum$/7.7~$\mum$ is about 10 times lower than that of neutral PAHs. Taken at face value, the 11.3~$\mum$/7.7~$\mum$ ratios observed in quasars indicate that the ionization fraction (or $U \sqrt{T_e}/n_e$) is about an order of magnitude higher compared to neutral PAHs. This scenario is possible if quasars participate in the heating of PAHs and in regulating their ionization fraction. If so, H atoms might be peeled off from the benzene ring, which would lower C-H/C-C and further reduce the 11.3~$\mum$/7.7~$\mum$ ratio. This process may be viable, in light of the lower PAH~8.6~$\mum$/7.7~$\mum$ ratios observed in quasars (Figure~\ref{fig:hist_pah_ratio}b), as the 8.6~$\mum$ band arises mainly from the C-H in-plane bending modes of neutral PAHs.  More data are needed to test this scenario, given the limited PAH~8.6~$\mum$ measurements currently available. 

The systematically low values of PAH~11.3~$\mum$/17.0~$\mum$ in quasars provide further corroborating evidence that powerful AGNs deplete small grains. On the basis of the observed suppressed PAH~11.3~$\mum$/17.0~$\mum$ ratios in strong AGNs, LaMassa \etal (2012) also suggested that the PAH carrier of the 11.3~$\mum$ feature is depleted. Although the chemical carrier of PAH~17.0~$\mum$ is still uncertain (\eg Boersma \etal 2010), it has been proposed that this feature originates from large ($N_{\rm c} \approx 10^{3}-10^{4}$) PAH molecules (\eg van Kerckhoven \etal 2000; Smith \etal 2007; Draine \etal 2021) that arise from the in-plane or out-of-plane bending mode of a C-C-C skeleton structure (van Kerckhoven \etal 2000; Peeters \etal 2004a).  The strength of PAH~17.0~$\mum$ is sensitive to the ionization parameter $U$: increasing $U$ from $10^{2}$ to $10^4$ lowers the PAH~17.0~$\mum$ strength by a factor of $\sim 2$ (Draine \etal 2021, their Figure~19).  When $U > 10^2$, larger PAH molecules cannot cool completely between single photon absorption events, gradually increasing their temperature so that they radiate at shorter wavelength bands. This effect elevates PAH~11.3~$\mum$/17.0~$\mum$. Two factors can produce a low value of PAH~11.3~$\mum$/17.0~$\mum$: a predominance of larger grains or a low $U$.  The second option seems highly unlikely. First, $75\%-90\%$ of the host galaxies of PG quasars are actively forming stars (Xie \etal 2021), and thus their value of $U$ cannot be lower than that of typical SFGs, let alone that of local low-luminosity AGNs, which have even lower levels of star formation activity (e.g., Ho et al. 2003; Ho 2008; Ellison \etal 2016). And second, the dust that accounts for the $5-8~\mum$ continuum in quasars has a temperature of $\sim 600$\, K (Xie et al. 2017), approximately twice as hot as in SFGs or low-luminosity AGNs ($\sim 300$\,K; Xie \etal 2018b). Given that the dust temperature scales as $T_{\rm d} \approx U^{1/6}$ (\eg Draine \& Li 2007), a factor of 2 increase in $T_{\rm d}$ results in $\sim 60$ times greater heating in quasars. Hence, it is highly unlikely that quasar host galaxies have lower $U$ than SFGs or low-luminosity AGNs. Instead, the systematically lower PAH~11.3~$\mum$/17.0~$\mum$ ratios in quasars most likely reflect a deficit of smaller grains.

\begin{figure}
\begin{center}
\includegraphics[width=\textwidth]{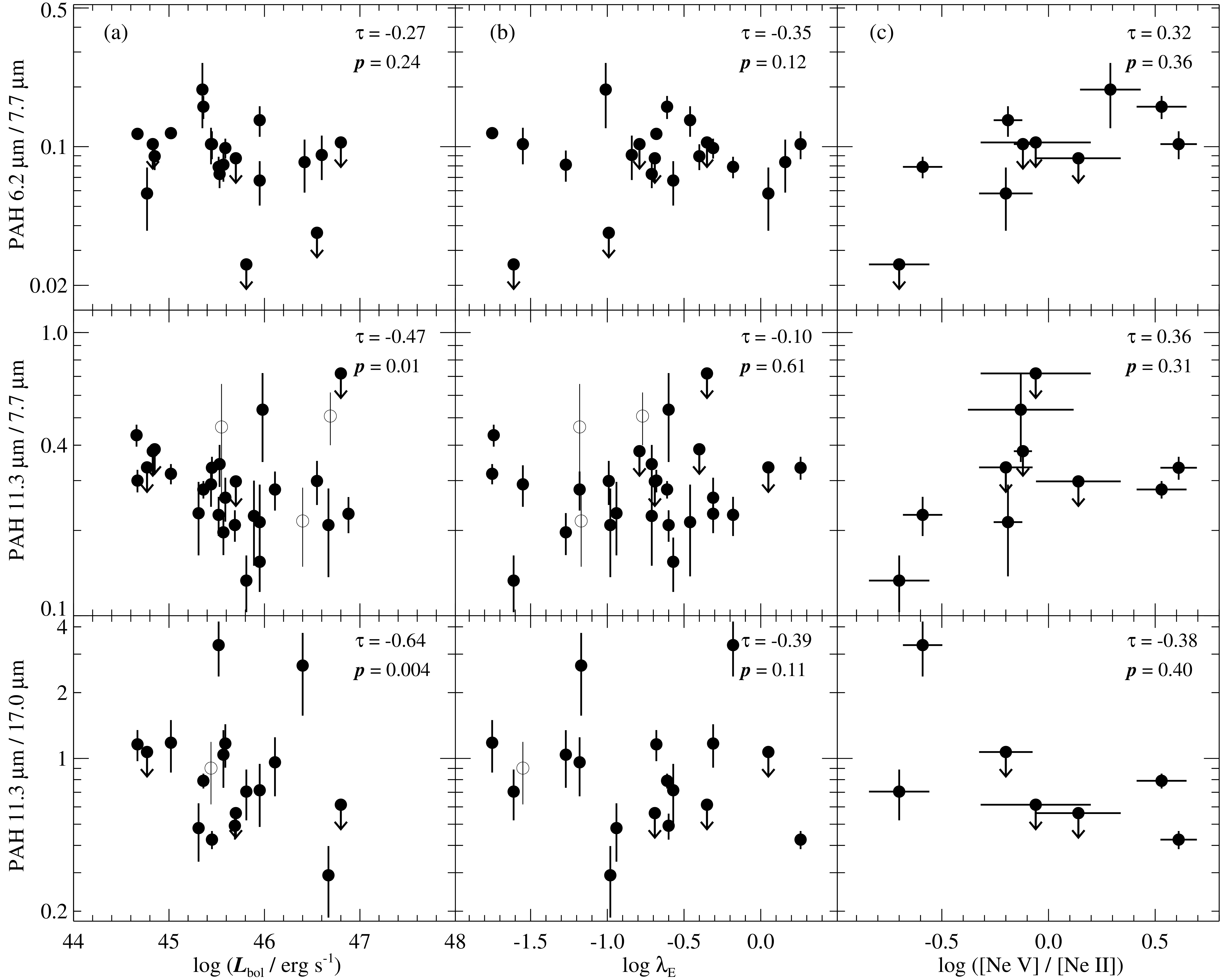}
\caption{Dependence of (top) PAH~$6.2\ \mum$/$7.7\ \mum$, (middle) PAH~11.3~$\mum$/7.7~$\mum$, and (bottom) PAH~11.3~$\mum$/17.0~$\mum$ on (a) \lbol, (b) $\lambda_{\mathrm{E}}$, and (c) $\rm [Ne~V]/[Ne~II]$.  The open circles indicate objects whose $\agnfracpg$ is less well-constrained from the IR SED fitting. The correlation coefficient $\tau$ and its corresponding $p$ value from {\tt linmix} are given in the top-right corner of each panel. We only include upper limits that are not affected by AGN dilution.}
\label{fig:pah_ratio_agn}
\end{center}
\end{figure}

The distinctive PAH properties of luminous AGNs suggest that the BH accretion influences the size distribution, ionization fraction, and chemistry of PAHs. In Figure~\ref{fig:pah_ratio_agn}, we look for possible trends between the PAH~6.2~$\mum$/7.7~$\mum$, 11.3~$\mum$/7.7~$\mum$, and PAH~11.3~$\mum$/17.0~$\mum$ ratios with the quasar bolometric luminosity, Eddington ratio, and [Ne~V]~14.32~$\mum$/[Ne~II]~12.81~$\mum$. As PG quasars are luminous AGNs, we adopt [Ne~V]/[Ne~II] instead of the more commonly used [Ne~III]/[Ne~II] to represent the hardness of the AGN radiation field to mitigate contamination from star formation (Ho \& Keto 2007).  None of the trends is statistically significant, except for the inverse correlations between PAH~11.3~$\mum$/7.7~$\mum$ and \lbol\ ($\tau = -0.47$, $p = 0.01$) and between PAH~11.3~$\mum$/17.0~$\mum$ and \lbol\ ($\tau = -0.64$, $p = 0.004$), which support the notion that more powerful quasars have more ionized PAH molecules and larger size distribution.  Future work should extend the dynamic range of the AGN parameters and substantially expand the sample size.

\subsection{ Potential Sources of Systematic Uncertainty }

Prominent silicate emission features around 9.7 and 18~$\mum$ are detected in 53 out of the 86 PG quasars, and their PAH strength is inevitably sensitive to the subtraction of the silicate component, especially for PAH 11.3~$\mum$ and 17.0~$\mum$, which located near the peak of silicate features. For the 53 objects with silicate emission, we test the effect of our fitting method on the final PAH measurements; the mid-IR spectrum of the rest is dominated by dust continuum emission, and their PAH strength is barely affected by the silicate fitting process. To calculate the absorption coefficient, we consider grain sizes distributed from 0.1 to 10~$\mum$ and six different chemical compositions of silicate dust. For a fixed grain size, different silicate dust peaks at a similar wavelength but differs in width and shape (\eg Figure 4b in Xie \etal 2014). Amorphous olivine and pyroxene deviate from the silicate profile of most objects by showing a unique profile (\eg Figure 3 in Xie \etal 2017). Once we identify the best-matching chemical composition, we fix it and then adjust the grain size. We find that in most cases, when the grain size is smaller than 0.8~$\mum$ or larger than 2~$\mum$, the best fit predicts a bluer or redder silicate peak than observed. We, therefore, set the limits of the grain size distribution to these values and then derive the PAH strength following the method in Section~2. Since only 13/53 sources have $\agnfracpg\ \leq 10$ ($\agnfrac^{c}$ of PAH 7.7~$\mum$) and 3 have $\agnfracpg\ \leq 5$ ($\agnfrac^{c}$ of PAH 6.2~$\mum$ and 11.3~$\mum$), most of the PAH strengths are upper limits affected by AGN dilution. Studying the range of PAH strengths produced by our experiments involving different choices of silicate emission models, and taking into account upper limits using survival analysis, we find that the median (16\%, 84\%) of PAH~6.2~$\mum$/7.7~$\mum$ and PAH~8.6~$\mum$/7.7~$\mum$ are hardly affected by the details of the silicate fitting process. The reason is that PAH 6.2, 7.7, and 8.6~$\mum$ are located relatively far from the 9.7~$\mum$ silicate emission feature. PAH 11.3~$\mum$ and 17.0~$\mum$ can be more strongly affected by silicate emission, at 9.7 and 18~$\mum$, respectively. However, we have verified that none of our statistical results on differences between the PG quasars and SFGs and low-luminosity AGNs is changed by the modeling details of silicate emission. The conclusions of the SFR, which is based on the integrated PAH~5--15~$\mum$ emission, are similarly unaffected.

We close with a comment on the possible influence of aperture effects on our conclusions, echoing a concern raised by Smith et al. (2007) and LaMassa et al. (2012).  As the PG quasars are relatively distant (median $z = 0.142$, which corresponds to $694$~Mpc), the IRS short-low (long-low) slit aperture size of $3{\farcs}7$ (10{\farcs}7) encompasses a physical scale of $\sim 12$ (36) kpc. The IRS spectra, therefore, are essentially globally averaged spectra, including not only the active nucleus but all the star-forming regions in the host galaxy.  While the similarity of the PAH~11.3~$\mum$/7.7~$\mum$ ratio of quasars to that of SFGs might suggest that the PAH emission is dominated by the large-scale contribution from the host galaxy, we note that within the same aperture, we simultaneously detecte depressed PAH~6.2~$\mum$/7.7~$\mum$, 8.6~$\mum$/7.7~$\mum$, and 11.3~$\mum$/17.0~$\mum$ ratios, from which we conclude that, despite the large aperture, the PAH~11.3~$\mum$/7.7~$\mum$ ratio is sensitive to conditions affected by the AGN.

\section {Summary \label{sec:conclusion}}

PAH emission serves as a useful indicator of star formation activity and a diagnostic of the physical conditions of the interstellar medium in galaxies.  However, their applicability to the host galaxies of AGNs is controversial in view of the ability of PAH molecules to survive the harsh radiation field of an active nucleus.  The situation in AGNs powerful enough to qualify as quasars is especially poorly understood.  We investigate the nature of PAH emission and its feasibility as a SFR predictor by analyzing the $5-40\,\mum$ Spitzer IRS spectra of 86 low-redshift ($z < 0.5$) PG quasars.  We apply the template-fitting method of Xie et al. (2018a) to decompose the PAH spectrum from the underlying continuum and silicate emission, which enables us to measure the flux of or place rigorous upper limits on the integrated ($5-15\,\mum$) PAH flux, as well as the individual bands at 6.2, 7.7, 8.6, 11.3, and 17.0~$\mum$.  To elucidate the origin of the weak PAH emission often observed in quasars, we generate mock spectra to mimic the effect of line dilution by the hot dust continuum of the torus, using realistic torus templates generated from broadband fits of the SEDs of the PG quasars themselves. These simulations allow us to establish the maximum critical threshold of AGN-to-host flux ratio below which reliable measurements of PAH emission can be made.  The principal conclusions of this study are:

\begin{itemize}

\item The torus does prevent PAH emission from being detected when the AGN continuum substantially outshines the host galaxy cold dust component.  However, given the observed AGN-to-host flux ratios of the PG sample, PAH emission is absent in some objects where it could have been detected, indicating that PAH emission can be intrinsically weak, even in host galaxies known to possess substantial interstellar medium and ongoing star formation.  The harsh environment around powerful AGNs evidently destroys PAH molecules.  

\item Using SFRs independently measured from the mid-IR fine-structure neon lines and the total IR luminosity, we demonstrate that PAH emission can probe ${\rm SFRs} \lesssim 50\,\usfr$, as typically found in quasars with $L_{\rm bol} \lesssim 10^{46}\, \rm erg\,s^{-1}$, but the systematically suppressed PAH emission in quasars with $L_{\rm bol} \gtrsim 10^{46}\, \rm erg\,s^{-1}$ leads to SFRs that are underestimated by $\sim 0.5$~dex.

\item On galactic scales, quasar host galaxies present lower 6.2~$\mum$/7.7~$\mum$, 8.6~$\mum$/7.7~$\mum$, and 11.3~$\mum$/17.0~$\mum$ but comparable 11.3~$\mum$/7.7~$\mum$ compared to star-forming galaxies and low-luminosity AGNs.  These trends are consistent with model predictions for larger and more ionized PAH molecules, suggesting that powerful AGNs preferentially destroy small grains and enhance the ionization fraction of the dust. 
 
\end{itemize}

\acknowledgments

We thank the referee for thoughtful and expert comments. This work was supported by the National Key R\&D Program of China (2016YFA0400702), the National Science Foundation of China (11721303, 11991052), and the National Natural Science Foundation of China for Youth Scientist Project (11803001).  Jinyi Shangguan and Ming-Yang Zhuang provided the torus templates and AGN fractions.  We are grateful to them and to Lulu Zhang for helpful comments on the manuscript.  We thank Yong Shi for sharing the IRS dataset of the PG quasars. The Combined Atlas of Sources with Spitzer IRS Spectra (CASSIS) is a product of the IRS instrument team, supported by NASA and JPL. CASSIS is supported by the ``Programme National de Physique Stellaire'' (PNPS) of CNRS/INSU co-funded by CEA and CNES and through the ``Programme National Physique et Chimie du Milieu Interstellaire'' (PCMI) of CNRS/INSU with INC/INP co-funded by CEA and CNES.  


\begin{deluxetable}{l c r c c l r c c c c c c r r r}
\rotate
\tablecaption{Physical Properties and Measurements of PG Quasars \label{tab:all}}
\tabletypesize{\scriptsize}
\tablehead{
\colhead{Object} &
\colhead{$z$} &
\colhead{$D_L$} &
\colhead{log $M_\mathrm{BH}$} &
\colhead{log $L_{\rm bol}$} &
\colhead{log $M_*$} &
\colhead{$f^{\rm PG}_{\rm AGN}$} &
\colhead{log $L^{\rm PAH}_{6.2 \mu\rm m}$} &
\colhead{log $L^{\rm PAH}_{7.7 \mu\rm m}$} &
\colhead{log $L^{\rm PAH}_{8.6 \mu\rm m}$} &
\colhead{log $L^{\rm PAH}_{11.3 \mu\rm m}$} &
\colhead{log $L^{\rm PAH}_{17.0 \mu\rm m}$} &
\colhead{log $L^{\rm PAH}_{5-15 \mu\rm m}$} &
\colhead{log $\rm SFR_{PAH}$}  & 
\colhead{log $\rm SFR_{Ne}$} &
\colhead{log $\rm SFR_{IR}$} \\
\colhead{} &
\colhead{} &
\colhead{(Mpc)} &
\colhead{($M_\odot$)} &
\colhead{(erg s$^{-1}$)} &
\colhead{($M_\odot$)} &
\colhead{} &
\colhead{(erg s$^{-1}$)} &
\colhead{(erg s$^{-1}$)} & 
\colhead{(erg s$^{-1}$)} &
\colhead{(erg s$^{-1}$)} &
\colhead{(erg s$^{-1}$)} &
\colhead{(erg s$^{-1}$)} &
\colhead{($M_\odot\ \rm yr^{-1}$)} &
\colhead{($M_\odot\ \rm yr^{-1}$)} &
\colhead{($M_\odot\ \rm yr^{-1}$)} \\ 
\colhead{(1)} &
\colhead{(2)} &
\colhead{(3)} &
\colhead{(4)} &
\colhead{(5)} &
\colhead{(6)} &
\colhead{(7)} &
\colhead{(8)} &
\colhead{(9)} &
\colhead{(10)} &
\colhead{(11)} & 
\colhead{(12)} & 
\colhead{(13)} & 
\colhead{(14)} & 
\colhead{(15)} &
\colhead{(16)}
}
\startdata
PG~0003$+$158 & 0.450 & 2572 &       9.45 &     46.99 &  11.83$^{\ast}$   &    \nodata  &                  $< 42.60 $  &                  $< 43.36 $  &                  $< 42.77 $  &                  $< 42.66 $ &                  $< 43.51 $  &                  $< 43.80 $  &                   $< 1.44 $  &                     \nodata  &                   $< 1.25 $ \\
PG~0003$+$199 & 0.025 &  113 &       7.52 &     45.17 &  10.38$^{\ast}$   &      36.0  &                  $< 40.97 $  &                  $< 41.86 $  &                  $< 41.57 $  &                  $< 41.84 $  &                  $< 42.30 $  &                  $< 42.50 $  &                   $< 0.21 $  &    $ 0.29_{-0.17}^{+0.17} $  &   $ -0.24_{-0.03}^{+0.03} $ \\
PG~0007$+$106 & 0.089 &  420 &       8.87 &     45.79 &           11.03   &      13.3  &                  $< 41.62 $  &   $ 42.81_{-0.07}^{+0.07} $  &                  $< 42.24 $  &                  $< 42.26 $  &   $ 42.36_{-0.10}^{+0.10} $  &   $ 43.20_{-0.05}^{+0.05} $  &    $ 0.87_{-0.06}^{+0.06} $  &    $ 1.19_{-0.15}^{+0.15} $  &    $ 0.92_{-0.03}^{+0.03} $ \\
PG~0026$+$129 & 0.142 &  693 &       8.12 &     46.07 &           11.07   &      28.4  &                  $< 42.08 $  &                  $< 42.32 $  &                  $< 40.76 $  &                  $< 42.00 $  &                  $< 42.55 $  &                  $< 42.90 $  &                   $< 0.59 $  &    $ 1.07_{-0.24}^{+0.24} $  &    $ 0.51_{-0.04}^{+0.03} $ \\
PG~0043$+$039 & 0.384 & 2133 &       9.28 &     46.51 &           11.13   &    \nodata  &                  $< 41.96 $  &   $ 42.68_{-0.08}^{+0.08} $  &                  $< 40.46 $  &                  $< 42.49 $ &                  $< 42.31 $  &                  $< 43.50 $  &                   $< 1.16 $  &                     \nodata  &                   $< 1.45 $ \\
PG~0049$+$171 & 0.064 &  297 &       8.45 &     44.97 &  11.07$^{\ast}$   &      36.0  &                  $< 40.08 $  &                  $< 41.80 $  &                  $< 41.46 $  &                  $< 41.42 $  &                  $< 41.61 $  &                  $< 42.20 $  &                  $< -0.10 $  &                     \nodata  &   $ -0.43_{-0.07}^{+0.05} $ \\
PG~0050$+$124 & 0.061 &  282 &       7.57 &     45.76 &           11.31   &      19.8  &                  $< 41.99 $  &                  $< 43.12 $  &                  $< 42.98 $  &                  $< 42.95 $  &                  $< 43.03 $  &                  $< 43.60 $  &                   $< 1.21 $  &                 [0.19, 0.92]  &    $ 1.60_{-0.01}^{+0.01} $ \\
PG~0052$+$251 & 0.155 &  763 &       8.99 &     46.00 &           11.24   &      16.2  &                  $< 42.34 $  &   $ 43.19_{-0.07}^{+0.07} $  &                  $< 42.44 $  &                  $< 42.69 $  &                  $< 42.94 $  &   $ 43.60_{-0.09}^{+0.09} $  &    $ 1.21_{-0.10}^{+0.10} $  &                     \nodata  &    $ 1.16_{-0.02}^{+0.02} $ \\
PG~0157$+$001 & 0.164 &  811 &       8.31 &     45.95 &           11.72   &       3.0  &   $ 42.96_{-0.05}^{+0.05} $  &   $ 43.82_{-0.06}^{+0.06} $  &                  $< 43.31 $  &   $ 43.15_{-0.14}^{+0.14} $  &                  $< 43.38 $  &   $ 44.30_{-0.04}^{+0.04} $  &    $ 1.90_{-0.06}^{+0.06} $  &    $ 2.38_{-0.14}^{+0.14} $  &    $ 2.50_{-0.05}^{+0.04} $ \\
PG~0804$+$761 & 0.100 &  475 &       8.55 &     46.03 &           10.83   &      36.0  &                  $< 41.64 $  &                  $< 42.98 $  &                  $< 42.65 $  &                  $< 42.73 $  &                  $< 42.97 $  &                  $< 43.60 $  &                   $< 1.20 $  &                 [0.76, 1.25]  &    $ 0.48_{-0.05}^{+0.07} $ \\
PG~0838$+$770 & 0.131 &  635 &       8.29 &     45.70 &           11.33   &       4.4  &                  $< 41.83 $  &   $ 42.89_{-0.05}^{+0.05} $  &                  $< 42.22 $  &                  $< 42.36 $  &   $ 42.61_{-0.10}^{+0.10} $  &   $ 43.10_{-0.12}^{+0.12} $  &    $ 0.76_{-0.13}^{+0.13} $  &    $ 0.63_{-0.26}^{+0.26} $  &    $ 1.37_{-0.04}^{+0.03} $ \\
PG~0844$+$349 & 0.064 &  297 &       8.03 &     45.46 &           10.88   &      23.4  &                  $< 40.97 $  &   $ 42.28_{-0.05}^{+0.05} $  &                  $< 41.65 $  &                  $< 41.81 $  &   $ 42.05_{-0.13}^{+0.13} $  &   $ 42.60_{-0.08}^{+0.08} $  &    $ 0.30_{-0.08}^{+0.08} $  &    $ 0.29_{-0.21}^{+0.21} $  &    $ 0.26_{-0.02}^{+0.02} $ \\
PG~0921$+$525 & 0.035 &  159 &       7.45 &     44.60 &  10.32$^{\ast}$   &      40.7  &                  $< 39.64 $  &   $ 41.38_{-0.07}^{+0.07} $  &                  $< 40.48 $  &                  $< 40.98 $  &                  $< 41.51 $  &                  $< 41.80 $  &                  $< -0.47 $  &    $ 0.57_{-0.15}^{+0.15} $  &   $ -0.26_{-0.02}^{+0.02} $ \\
PG~0923$+$201 & 0.190 &  955 &       9.33 &     46.01 &           11.28   &      51.6  &                  $< 41.61 $  &                  $< 42.66 $  &                  $< 42.42 $  &                  $< 42.49 $  &                  $< 42.84 $  &                  $< 43.30 $  &                   $< 0.95 $  &                 [0.62, 1.25]  &    $ 0.64_{-0.05}^{+0.05} $ \\
PG~0923$+$129 & 0.029 &  131 &       7.52 &     44.83 &  10.38$^{\ast}$   &       3.7  &                  $< 41.27 $  &   $ 42.25_{-0.08}^{+0.08} $  &                  $< 41.76 $  &                  $< 41.83 $  &                  $< 42.04 $  &   $ 42.70_{-0.06}^{+0.06} $  &    $ 0.35_{-0.07}^{+0.07} $  &    $ 0.56_{-0.15}^{+0.15} $  &    $ 0.70_{-0.02}^{+0.01} $ \\
PG~0934$+$013 & 0.050 &  229 &       7.15 &     44.85 &  10.10$^{\ast}$   &       3.5  &   $ 41.26_{-0.05}^{+0.05} $  &   $ 42.30_{-0.04}^{+0.04} $  &                  $< 41.54 $  &                  $< 41.89 $  &                  $< 42.15 $  &   $ 42.70_{-0.07}^{+0.07} $  &    $ 0.37_{-0.08}^{+0.08} $  &                     \nodata  &    $ 0.61_{-0.02}^{+0.02} $ \\
PG~0947$+$396 & 0.206 & 1045 &       8.81 &     45.78 &           10.92   &      39.0  &                  $< 41.64 $  &   $ 42.78_{-0.11}^{+0.11} $  &                  $< 42.43 $  &                  $< 42.52 $  &                  $< 42.95 $  &                  $< 43.20 $  &                   $< 0.88 $  &                     \nodata  &    $ 0.94_{-0.04}^{+0.05} $ \\
PG~0953$+$414 & 0.239 & 1235 &       8.74 &     46.35 &           11.35   &      46.6  &                  $< 42.45 $  &                  $< 43.03 $  &                  $< 42.45 $  &                  $< 42.89 $  &                  $< 42.94 $  &                  $< 43.40 $  &                   $< 1.07 $  &    $ 1.29_{-0.24}^{+0.24} $  &    $ 0.95_{-0.07}^{+0.07} $ \\
\enddata
\tablecomments{
Col. (1): Object name.
Col. (2): Redshift.
Col. (3): Luminosity distance.
Col. (4): Mass of the BH, from Shangguan et al. (2018).
Col. (5): Bolometric luminosity of the AGN, from Shangguan et al. (2018).
Col. (6): Stellar mass of the quasar host galaxy, from Zhang \etal (2016) or derived from the $M_{\mathrm{BH}}-M_{\ast}$ relation of Greene et al. (2020; marked with an asterisk).
Col. (7): AGN fraction.
Col. (8): PAH luminosity at 6.2 $\mu$m. 
Col. (9): PAH luminosity at 7.7 $\mu$m.
Col. (10): PAH luminosity at 8.6 $\mu$m.
Col. (11): PAH luminosity at 11.3 $\mu$m.
Col. (12): PAH luminosity at 17.0 $\mu$m.
Col. (13): PAH luminosity within 5--15 $\mu$m.
Col. (14): SFR derived from the PAH luminosity within 5--15 $\mu$m. 
Col. (15): SFR derived from the neon line luminosity.
Col. (16): SFR derived from the total IR luminosity.
(Table~1 is published in its entirety in machine-readable format. A portion is shown here for guidance regarding its form and content.)
}
\end{deluxetable}


\end{document}